\documentclass[10pt,twocolumn,letterpaper]{article}
\usepackage[letterpaper,
  textwidth=7in,
  textheight=9in,
  top=1in,
  headheight=0pt,
  headsep=0pt,
  columnsep=0.33in,
  left=0.75in,
  includefoot]{geometry}
\usepackage[T1]{fontenc}
\usepackage[utf8]{inputenc}
\usepackage{newtxtext,newtxmath}
\usepackage[scaled=0.88]{beramono}
\usepackage{microtype}
\usepackage{booktabs}
\usepackage{array}
\usepackage{tabularx}
\usepackage{adjustbox}
\usepackage{float}
\usepackage{stfloats}

\usepackage{longtable}
\usepackage{fancyvrb}
\usepackage[hidelinks,breaklinks=true]{hyperref}
\usepackage{xcolor}
\usepackage{enumitem}
\usepackage{balance}
\usepackage{tikz}
\usetikzlibrary{arrows.meta,positioning,calc}
\setlist[itemize]{noitemsep,topsep=1pt,partopsep=0pt,parsep=0pt,leftmargin=*}
\setlist[enumerate]{noitemsep,topsep=1pt,partopsep=0pt,parsep=0pt,leftmargin=*}
\usepackage{titlesec}
\titleformat{\section}{\normalsize\bfseries}{§\thesection}{0.6em}{}
\titleformat{\subsection}{\small\bfseries}{\thesubsection}{0.6em}{}
\titleformat{\subsubsection}{\small\bfseries}{\thesubsubsection}{0.6em}{}
\titleformat{\paragraph}{\small\bfseries}{\theparagraph}{0.6em}{}
\titlespacing*{\section}{0pt}{6pt plus 1pt minus 1pt}{2pt}
\titlespacing*{\subsection}{0pt}{4pt plus 1pt minus 1pt}{1pt}
\titlespacing*{\subsubsection}{0pt}{4pt plus 1pt minus 1pt}{1pt}
\titlespacing*{\paragraph}{0pt}{3pt plus 0.5pt minus 0.5pt}{0.5em}
\newcolumntype{Y}{>{\raggedright\arraybackslash}X}

\newcommand{\secref}[1]{§\ref{#1}}

\newcommand{\appref}[1]{App.~\ref{#1}}
\newcommand{\Appref}[1]{Appendix~\ref{#1}}

\title{\textbf{AGNT2: Autonomous Agent Economies on Interaction-Optimized Layer 2 Infrastructure}\footnotemark[1]\vspace{-0.5em}}
\author{\small Anbang Ruan \quad Xing Zhang \\
\small NetX Foundation \\
\small \texttt{ruan@netx.foundation} \quad \texttt{xing.zhang@netx.world}}
\date{}

\begin{document}
\twocolumn[
  \begin{@twocolumnfalse}
    \maketitle
  \end{@twocolumnfalse}
]
\thispagestyle{plain}
\begingroup
\renewcommand{\thefootnote}{\fnsymbol{footnote}}
\footnotetext[1]{Submitted for peer review conference consideration.}
\endgroup

\begin{abstract}

Current blockchain Layer 2 solutions are optimized for human-initiated financial transactions such as token transfers, swaps, and lending positions. Autonomous AI agents instead generate high-frequency, semantically rich, machine-to-machine service invocations that existing general-purpose chains do not natively model. We present AGNT2, an agent-native Layer 2 system architecture implemented as a three-layer stack for inter-agent and microservice on-chain communication. AGNT2 combines three elements: a sidecar deployment pattern that makes existing Dockerized services addressable as on-chain agents without application-logic code modification; a three-layer architecture with P2P state channels for established bilateral pairs (Layer Top, <100 ms), a dependency-aware execution layer for first-contact and multi-party interactions (Layer Core, 500 ms--2 s, 300K--500K TPS design target), and a settlement layer with computational fraud proofs anchored to any EVM L1 (Layer Root); and an agent-native execution environment plus interaction trie that make service invocation, identity, reputation, capabilities, and session context first-class protocol objects. The paper focuses on execution-layer systems questions: sequencing, state, settlement, and the data-availability (DA) bandwidth limit. AGNT2 targets 300K--500K TPS under optimistic assumptions, but this remains a design target: prototype measurements validate selected components, while simulation and analytical modeling support the broader design. Current DA backends likely cap practical deployment near 10K--100K TPS, leaving a roughly 100$\times$ gap at the target ceiling. Our thesis is that, for the open setting AGNT2 targets, agent-native execution semantics fit more cleanly in a dedicated execution layer than as an application atop a general-purpose chain.

\end{abstract}

\section{Introduction}\label{sec:introduction}

A content-pipeline agent calls transcription services, passes the transcript on for translation, and pays each party directly. Similar patterns already appear in trading and DevOps, where agents negotiate with external services, coordinate specialized subagents, hold balances, and may accrue revenue on their own account. Historically, the transaction boundary assumed a human-controlled EOA (externally owned account): a key was controlled, a person signed, and only then did tokens move or positions change. Today, smart accounts, delegated signers, session keys, and agent wallets have already weakened that assumption [36, 37]. Software can act within delegated limits without per-step human approval. That is real progress, but it mostly solves authorization, not execution semantics. To the chain, these agent behaviors still appear as generic calls and token transfers rather than first-class service invocations with response delivery, dependency ordering, capability discovery, and portable interaction reputation. AGNT2 begins from that mismatch.

In the open, cross-principal setting AGNT2 targets, coordination needs more than autonomous payment. It also needs portable identity, reputation that third parties can inspect, settlement that spans multi-agent composition graphs, and dispute handling that does not depend on bespoke bilateral preconditions. Emerging standards already cover parts of this bundle. The proposed ERC-8004 provides interfaces for agent identity, reputation, and validation, while x402 supports programmatic agent-to-agent payments for API access [38, 39]. What remains harder is composing these pieces into open multi-agent workflows with native dependency ordering and atomic cross-service settlement, without reintroducing a trusted intermediary. In closed deployments, parts of this bundle can be approximated off-chain; the difficulty is specific to the permissionless setting. Because LLM-driven autonomous agents and tool ecosystems are already in production [18, 19, 24], this transition is not hypothetical. With the sidecar pattern, existing Dockerized services can already be surfaced as on-chain agents with no application-logic code change. Timing therefore matters: if HTTP plus centralized rails become the default before purpose-built infrastructure exists, retrofitting trustlessness later will be materially harder and costlier. \secref{subsec:why-onchain} develops this argument against the strongest off-chain alternative, namely TEE-attested logs with SLA arbitration.

What current stacks still lack is a native execution substrate for those interactions. Existing rollups accept generic calldata and expose generic contract calls; they do not treat an agent call as a typed service invocation with payment escrow, native agent identity, or dependency metadata the sequencer can schedule around. Rich payloads are still priced as calldata, challenge windows inherit human-financial assumptions, and session context, capability registries, and interaction history live above the protocol rather than in its state model. This missing substrate is the gap AGNT2 is meant to close. We therefore make service invocation, rather than token transfer, the basic unit of computation, and our architecture follows from that choice.

AGNT2 is an agent-native Layer 2 system architecture implemented as a three-layer stack above an EVM-compatible L1. \secref{sec:architecture} unpacks the roles of Layer Root as settlement, Layer Core as the agent-optimized execution layer, and Layer Top as the bilateral fast path.

\begin{enumerate}
  \item \textbf{The Sidecar Pattern} (\secref{subsec:sidecar-pattern}) --- Existing Dockerized services become on-chain agents by injecting an \texttt{\small agnt2/sidecar} container alongside them. Zero application-logic code change. The sidecar handles chain listening, identity, payment escrow, attestation, and DA (data availability) posting.
  \item \textbf{The Three-Layer L2 Stack} (\secref{sec:architecture}) --- Layer Top (<100 ms P2P, off-sequencer bilateral throughput bounded by endpoint resources), Layer Core (500 ms--2 s, 300K--500K TPS design target, dependency-aware), Layer Root (settlement and Type 1 computational fraud proofs anchored to any EVM L1). Roughly four orders of magnitude in latency from Layer Top to L1 finality (\textasciitilde{}100 ms vs. \textasciitilde{}12 min) and two to three orders of magnitude in per-interaction cost.
  \item \textbf{The Agent-Native Execution Environment} (\secref{subsec:layer-core}) --- A custom VM with INVOKE, RESPOND, COMPOSE, and DISCOVER as first-class opcodes plus capability-weighted gas metering.
  \item \textbf{The Interaction Trie} (\secref{subsec:layer-core}) --- A purpose-built Merkle-ized state structure encoding agent identity (proposed ERC-8004 compatible), reputation, capabilities, channels, and session context as first-class protocol objects.

\end{enumerate}
Existing rollups and agent platforms can be compared today; production throughput under real agent workloads cannot, because no deployment exists yet. That is the Frontiers track split: what we can test now, and what has to wait. We evaluate design-space differentiation against existing rollups and agent platforms (\secref{sec:background}, \secref{sec:architecture}), analytical throughput bounds, including the \textasciitilde{}100$\times$ DA bandwidth gap that operationally governs them (\secref{sec:evaluation}, \appref{app:throughput-analysis}), the execution-layer threat model (\secref{sec:security}, \appref{app:extended-threat-model}), the argument for why the open setting benefits from an on-chain substrate (\secref{subsec:why-onchain}), and the argument for why typed agent interactions motivate a native execution layer rather than a pure app-layer construction (\secref{subsec:why-new-execution-layer}). We defer cryptoeconomic incentive-compatibility (\secref{subsubsec:cryptoeconomic-design}) and on-chain governance of semantic fraud (\secref{subsubsec:qualitative-adjudication}). The narrowing is intentional. It keeps our analytical results defensible within a single paper.

Appendices A--G carry the deferred technical detail: transaction grammar (\appref{app:transaction-grammar}), sidecar failure handling (\appref{app:sidecar-failure}), fraud-proof derivations (\appref{app:fraud-proof-derivations}), interaction-trie schema (\appref{app:interaction-trie}), extended throughput derivations (\appref{app:throughput-analysis}), extended threat model (\appref{app:extended-threat-model}), and extended related work (\appref{app:extended-related-work}). We use \secref{sec:background} for background, related work, and design positioning, closing it with the on-chain/open-setting argument (\secref{subsec:why-onchain}) and the native-execution-layer argument (\secref{subsec:why-new-execution-layer}). \secref{sec:problem-setting} then fixes the deployment setting and derives design requirements; \secref{sec:architecture} gives the architecture, including the sidecar pattern, three-tier stack, agent-native VM, and interaction trie. \secref{sec:security} states the threat model. \secref{sec:evaluation} evaluates requirements satisfaction, throughput characteristics, and security coverage, while \secref{sec:implementation} records implementation status. \secref{sec:discussion} discusses limitations, tradeoffs, assumptions, and deferred research agendas. \secref{sec:conclusion} concludes.

\section{Background and Design Positioning}\label{sec:background}

\subsection{Existing Approaches and Closest Precursors}\label{subsec:closest-precursors}

The closest precursors fall into five buckets: rollups, agent systems on blockchain, agent communication protocols, microservice infrastructure, and dependency-aware execution engines. Taken together, they cover many adjacent pieces of the problem, but none combines permissionless service invocation, escrowed interaction semantics, dependency-aware sequencing, and agent-native state in one execution layer.

\textbf{Rollups.} Optimism and Arbitrum show one Layer 2 family: \textit{optimistic rollups} execute off-chain against an EVM, post compressed transactions and state roots to L1, and rely on challenge-window fraud proofs [1, 12, 13]. In another family, \textit{ZK rollups} get immediate L1 finality from batch validity proofs. The cost is proof generation, and EVM-equivalence still constrains the instruction set [3]. \textit{App-specific rollups} come closer to AGNT2 because they allow custom execution environments, but they remain general-purpose environments with tunable parameters [12, 28]. All three target EVM-compatible financial transactions rather than service invocation with escrowed payment, caller/callee asymmetry, dependency awareness, or identity, reputation, and session state.

\textbf{Agent systems on blockchain.} Several projects combine AI agents and blockchain, but none redesigns the chain for agent-native interaction. Autonolas (Olas) coordinates off-chain agents through an on-chain service registry [20]. Fetch.ai is a Cosmos-based chain with agent messaging, predating LLM agents and oriented toward IoT and search [21, 28]. Ritual Network targets verifiable on-chain ML inference, not agent coordination [29]. Phala Network uses TEEs for off-chain agent execution with on-chain integrity verification, but lacks a general-purpose multi-agent interaction layer [22].

Two adjacent families require a "why-not-this-instead" answer. \textit{Hyperledger Fabric} provides consortium identity and channels but assumes known ratifying organizations; AGNT2 targets a permissionless protocol [27]. \textit{Cosmos application-specific zones} would need custom SDK modules for AGNT2's dependency-aware sequencing, agent-typed VM, and interaction trie, which means building AGNT2 \textit{on} Cosmos rather than as an EVM rollup; AGNT2's optimistic-rollup fraud-proof lineage also amortizes batch cost in a way Tendermint BFT does not at 500K TPS [28]. Appendix G.1 gives per-project details. The common gap is that existing projects treat the chain as a registry/settlement layer for off-chain interactions, or assume governance boundaries absent in the open agent economy. None makes permissionless agent service invocation --- request, response, payment, attestation --- a first-class on-chain primitive at the target throughput.

\textbf{Agent communication protocols.} Anthropic's Model Context Protocol (MCP) is single-agent: one model, one tool set, one context, with no inter-agent semantics, payment, or on-chain record [18]. Google's Agent-to-Agent (A2A) protocol defines structured caller/callee/capability/result interactions over HTTP; it is structurally similar to AGNT2's on-chain primitive but lacks payment, escrow, reputation, dispute resolution, and permanent record [19]. A2A is the interaction protocol AGNT2 takes on-chain. Classical FIPA-ACL and KQML addressed message-passing grammar for trusted agent environments without economic or settlement semantics [30, 31].

\textbf{Microservice infrastructure.} The closest non-blockchain analog is the service mesh (Kubernetes, Istio, Envoy): infrastructure for discovery, load balancing, observability, and security for microservices [32, 33, 34]. Service mesh mechanisms (mTLS, liveness probes, circuit breakers, schema validation, authz, distributed tracing) work under a single trusted operator because SLAs and employment contracts enforce correctness. This model fails when services belong to \textbf{different principals with conflicting economic interests}: a mesh cannot verify result correctness, establish cross-interaction reputation, impose penalties, or support third-party audit. AGNT2 is to cross-principal agent interactions what a service mesh is to same-operator microservices, with no assumed trust and trustless economic settlement.

\textbf{Dependency-aware execution.} ROCOCO and Vegeta are the closest mechanistic precedents: DAG-based parallel transaction execution [7, 8]. ROCOCO uses client-declared access sets; Vegeta speculatively executes EVM contracts and builds conflict graphs from read-write intersections, reaching 7.8$\times$ speedup on DeFi workloads. AGNT2's Dependency Analyzer follows this lineage but uses protocol-declared dependencies: every RESPOND names its parent INVOKE by hash, and every COMPOSE names its constituents, so the executor need not discover dependencies at runtime. This narrowing yields O(n) DAG construction (vs. Vegeta's O(n²), infeasible at 250K tx/block), makes dependency order a \textit{correctness} invariant, and supports cryptographic attestation among mutually-untrusting principals. Appendix G.2 gives detailed positioning.

\subsection{Why On-Chain at All?}\label{subsec:why-onchain}

Could the agent economy use off-chain infrastructure --- Ray clusters with TEE attestation, append-only signed logs, and SLA arbitration [22, 35] --- or a permissioned distributed ledger (Hyperledger Fabric, Cosmos application zones; \secref{subsec:closest-precursors}) --- to replicate enough of AGNT2's semantics for the open setting we target? Our claim is narrower than "nothing off-chain can work": in closed or bilateral deployments, many components can. The question is whether the full bundle of \textbf{trust-minimized, third-party-verifiable multi-agent coordination without prior bilateral setup} can be reproduced without reintroducing a trusted intermediary. Three sub-properties decompose this claim; the strongest off-chain alternative weakens each.

\textbf{(a) Reputation without a trusted aggregator.} A minimally Byzantine attestor trivially breaks single-operator designs. A \textit{rotating cross-organizational TEE committee} (Appendix G.3) improves this to committee-level integrity, but not aggregator-free portability: a new verifier still has to decide whether an unfamiliar committee and its rotation history are legitimate, which requires out-of-band bootstrapping. In an open agent economy with previously unknown organizations, that verification burden grows with the number of committee lineages a newcomer must trust. On-chain reputation derivable from the state root reduces this to a single public verification target: any party can verify a Merkle path against a publicly auditable root without any prior relationship with the organizations whose agents contributed to that root.

\textbf{(b) Atomic multi-agent payment coordination.} A COMPOSE spanning five specialized agents where all succeed or all revert can be approximated off-chain if participants accept prior trust, bespoke escrow agreements, or a coordinating intermediary. AGNT2's claim is narrower: in the open setting, all-or-nothing settlement without prior bilateral trust still requires a public coordinator. Implementing that coordinator on a general-purpose L1 reconstructs AGNT2 Layer Core, but without solving the cost problem --- it pays both L1 gas overhead (21K--50K gas per EVM transaction) and the encoding cost of storing interaction state in EVM contract storage, precisely the costs \secref{subsec:why-new-execution-layer} quantifies as the motivation for a purpose-built execution layer. The reconstruction is not merely redundant; it incurs the full penalty of the EVM-hosted path that AGNT2 is designed to avoid, while still lacking native interaction-trie state for reputation, channel balances, and capability registry.

\textbf{(c) Dispute escalation without bilateral setup.} Within this paper's scope, the relevant property is not fully solved qualitative adjudication --- which we defer to \secref{subsubsec:qualitative-adjudication} --- but the ability to escalate execution and settlement disputes without pre-negotiated bilateral agreements. Off-chain SLA arbitration depends on bilateral preconditions (shared SLA, trusted arbitrator, enforceable ruling) that do not hold for unknown cross-organization counterparties. In an open agent economy where any agent may invoke any other, the number of bilateral relationships that may need prior agreement grows combinatorially with ecosystem size. AGNT2 instead provides a public, stake-backed path for computational disputes over state transitions; qualitative disputes remain future work, but the escalation path itself is no longer bilateral.

For the open, permissionless setting AGNT2 targets, these three properties are difficult to assemble from off-chain primitives without reintroducing either a trusted intermediary or substantial prior coordination. Table~\ref{tab:comparison} summarizes the operational contrast between the human-transaction model these systems largely inherit and the agent-interaction model AGNT2 targets.

\begin{table*}[!t]
\centering
\footnotesize
\caption{Human-initiated transactions vs. agent service interactions across nine operational dimensions.}
\label{tab:comparison}
\begin{tabularx}{\textwidth}{@{}YYY@{}}
\toprule
Dimension & Human Transactions & Agent Interactions \\
\midrule
Frequency & Seconds to minutes & Milliseconds to seconds \\
Payload structure & Simple value transfer (20--200 bytes) & Structured capability invocation (200 bytes--10 KB+) \\
Composition depth & 2--5 contract hops & 10--100+ chained calls per task \\
Latency tolerance & Minutes acceptable & Sub-second required for coordination \\
Identity model & EOA / smart wallet & Capability-bearing agent with reputation \\
Payment granularity & Token-level transfers & Per-call micropayments (0.001--0.01 token) \\
State requirements & Account balances & Agent state machines, sessions, context \\
Dispute nature & Financial (wrong amount) & Service quality (garbage output) \\
Error model & Transaction reverts & Timeout, partial completion, degraded quality \\
\bottomrule
\end{tabularx}
\end{table*}

\subsection{Why a New Execution Layer?}\label{subsec:why-new-execution-layer}

The strongest composability objection is that AGNT2 resembles an app-specific rollup: Arbitrum Orbit or OP Stack with custom precompiles, a typed agent registry contract, a DAG-aware sequencer policy, and bilateral state channels. Much of the stack \textit{could} be approximated that way, especially for an MVP or lower-throughput deployment. The narrower question for this paper is different: can AGNT2's typed service-invocation semantics remain an application layer while still preserving efficient fraud-proof re-execution and dependency-order correctness at the scale we target?

The first mismatch is \textbf{fraud-proof re-execution}. If INVOKE, RESPOND, and COMPOSE sit above the EVM, a disputed transition is still logically an AGNT2 transition, not a generic EVM transition; validity depends on interaction state, escrow state, reputation counters, and dependency metadata. Verifiability then requires re-encoding the AGNT2 transition system inside contracts or verifier precompiles. That route may be acceptable for a narrower system, but it does not eliminate AGNT2's complexity; it pushes it into EVM-hosted verification while paying for both the agent-typed logic and the enclosing EVM substrate.

The interaction trie has the same failure mode. In AGNT2, it is native to the Layer Root commitment scheme; INVOKE, RESPOND, and COMPOSE are Merkle-committed typed records. Fraud-proof bisection then narrows the dispute to one AGNT2 VM step. If we encode the same objects above EVM storage, each disputed step has to verify trie traversal, object decoding, and the state-machine update in Solidity. The verifier cost then scales with emulating the typed object model inside the EVM, not with one native AGNT2 step. Our necessity claim is therefore narrower than "the functionality is impossible above EVM": it is that efficient fraud-proofing for typed service-invocation state becomes materially less attractive when the entire object model is hosted one layer up.

A second gap separates AGNT2 from a Vegeta [8] + Orbit [12] composition: \textbf{dependency declaration as protocol semantics, not runtime discovery}. Vegeta \textit{discovers} read-write conflicts at runtime via storage-access analysis, treating reordering as a performance loss [8]. AGNT2 \textit{requires} every RESPOND to name its parent INVOKE by hash (\appref{app:executor-state-transitions}), so dependency-order violations become block-validity errors rather than performance anomalies. A custom sequencer could ingest those declarations, but if the typed interaction objects remain contract-level artifacts, the rollup still sees generic calldata at the execution and fraud-proof layers. Put differently: sequencer customization can recover some scheduling behavior, but not by itself the native typed state, gas model, or proof object that AGNT2's design assumes.

The terminal proof object and verifier cost comparison appears in \appref{app:fraud-proof-bisection}; in brief, we estimate 1.7--4.3$\times$ smaller witness bytes and 4--6$\times$ cheaper verifier gas for the native path (design projections; see footnotes there).

We therefore treat AGNT2 as an \textit{agent-typed execution substrate}, not merely a rollup configuration. Service invocation and dependency ordering are first-class VM objects; so are payment escrow and dispute resolution. The claim is not that no Orbit/OP-Stack-derived implementation is possible, but that once typed service-invocation state and fraud-proof semantics are central, a native execution layer is the cleaner and more scalable design point.

\section{Problem Setting and Requirements}\label{sec:problem-setting}

A Dockerized content intelligence deployment can consist of eight independently operated containerized services: web crawler, language detector, competing translation services, topic classifier, entity extractor, quality scorer, LLM summarizer, and distribution. No central administrator owns them. Operators bring services up and take them down as needed. There is no standing coordination layer. Given the design position in \secref{sec:background}, we now fix that deployment setting and derive the execution-layer requirements it imposes. In this setting, autonomous agents need to locate capabilities across organizations by reputation and price, then call them with atomically escrowed payment guaranteed iff service is delivered. Settling high-frequency bilateral relationships happens through periodic checkpoints. Version changes add another wrinkle: reputation must survive upgrades. Multi-service pipelines also have to compose atomically, with every stage reverting on any failure. Containerized microservices dominate deployment; Dockerized LLM agents are already in production [18, 19, 24, 32].

Sections 1 and 2 narrow the gap already: open agent workflows need more than autonomous payment, and existing stacks still approximate the rest through trust assumptions, bespoke bilateral setup, or app-layer composition. For the deployment above, the execution layer still lacks three things at once: typed service invocation, native coordination state, and dependency-aware settlement. Rich payloads remain expensive, identity, reputation, channels, and session context live above the protocol, and multi-stage service composition is not a first-class settlement object. The requirements below make those missing properties explicit.

Seven requirements emerge from this deployment setting and infrastructure gap. R1--R3 address the sequencing, latency, and payload problems. R4--R5 address identity and micropayment economics. R6--R7 cover atomic composition and verifiable correctness:

\begin{enumerate}[label=\textbf{R\arabic*:}, leftmargin=*, itemsep=2pt, topsep=3pt]
\item \textbf{Sub-second interaction latency.} Established bilateral pairs should complete through the Layer Top P2P fast path in under 100 ms, while new or multi-party interactions should complete through the Layer Core sequenced path in under 2 seconds.
\item \textbf{Rich payload support.} Structured inputs and outputs, capability schemas, session context, and attestation headers should be first-class primitives rather than prohibitively expensive calldata.
\item \textbf{Dependency-aware ordering.} The sequencer should extract the INVOKE/RESPOND dependency graph so dependent interactions batch atomically while independent interactions parallelize.
\item \textbf{Agent identity as first-class state.} Each agent should have protocol-native state for identity, capabilities, reputation, economics, and interaction history, updated by typed transitions rather than arbitrary contract calls.
\item \textbf{Micropayment channels.} State channels should settle thousands of interactions per on-chain transaction so per-interaction cost falls by multiple orders of magnitude.
\item \textbf{Composable atomic transactions.} A COMPOSE primitive should specify a dependency-ordered set of agent invocations that executes atomically, with all stages succeeding or all reverting.
\item \textbf{Computational fraud proof.} Incorrect state transitions should be disprovable by algorithmic re-execution on L1 within a bounded challenge window; qualitative fraud remains deferred to \secref{subsubsec:qualitative-adjudication}.
\end{enumerate}

\section{Architecture AGNT2}\label{sec:architecture}

\subsection{System Overview}\label{subsec:system-overview}

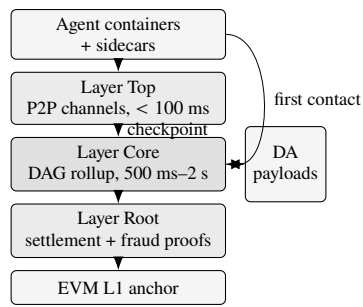
\begin{figure}[H]
\centering
\begin{tikzpicture}[
  font=\scriptsize,
  box/.style={draw, rounded corners=2pt, align=center, inner sep=3pt,
              minimum width=2.85cm, minimum height=0.45cm},
  side/.style={draw, rounded corners=2pt, align=center, inner sep=3pt,
              minimum width=1.05cm, minimum height=1.0cm},
  arrow/.style={-{Latex[length=1.8mm]}, line width=0.35pt}
]
\node[box, fill=black!4] (agent) {Agent containers\\+ sidecars};
\node[box, fill=black!8, below=0.16cm of agent] (top) {Layer Top\\P2P channels, $<100$ ms};
\node[box, fill=black!12, below=0.16cm of top] (core) {Layer Core\\DAG rollup, 500 ms--2 s};
\node[box, fill=black!8, below=0.16cm of core] (root) {Layer Root\\settlement + fraud proofs};
\node[box, fill=black!4, below=0.16cm of root] (l1) {EVM L1 anchor};
\node[side, fill=black!4, right=0.25cm of core] (da) {DA\\payloads};
\draw[arrow] (agent) -- (top);
\draw[arrow] (agent.east) .. controls +(0.65,0) and +(0.65,0) .. node[right, xshift=0.03cm] {first contact} (core.east);
\draw[arrow] (top) -- node[right] {checkpoint} (core);
\draw[arrow] (core) -- (root);
\draw[arrow] (root) -- (l1);
\draw[arrow] (core) -- (da);
\end{tikzpicture}
\caption{AGNT2. Each interaction picks the cheapest layer that'll still satisfy the required trust and ordering guarantees.}
\label{fig:architecture}
\end{figure}

A service doesn't change its own code. AGNT2 adds a \textbf{sidecar} (\texttt{\small agnt2/sidecar}): a lightweight container that runs beside it and handles the on-chain interface [33, 34], registers the service, advertises capabilities, manages payment escrow, and issues cryptographic attestation. Above this, the three-tier stack routes each interaction to the minimum-overhead mechanism matching its trust requirements (Figure~\ref{fig:architecture}). Established pairs take the P2P fast path. <100 ms latency, bilateral throughput bounded by endpoint resources. For verifiable first-contact and multi-party interactions, we use a sequenced rollup with 500 ms--2 s latency and a 300K--500K TPS design target; current DA backends constrain practical deployments to 10K--100K TPS (\secref{subsec:da-bandwidth-gap}, \secref{subsec:layer-core-execution}). Settlement batches state roots to an external L1 chain and handles computational fraud proofs for AGNT2 [1, 12, 13].

Three-speed routing reflects a quantitative reality. Agent interactions span roughly six orders of magnitude in frequency and latency tolerance. Pricing oracles need sub-100 ms round-trips and near-zero cost. first-contact multi-agent composition needs 500 ms--2 s ordering, escrowed atomic payment, and a verifiable record. L1 finality doesn't need those. And periodic anchoring suffices.

AGNT2 settles on its own Root layer, not directly on the L1. The L1 remains an external anchor, while Layer Root handles batching and fraud proofs; \secref{subsec:layer-root} explains that settlement path in detail.

\subsection{The Sidecar Pattern}\label{subsec:sidecar-pattern}

A Docker container, whether a LLM inference service, data pipeline, or pricing oracle, becomes an on-chain agent when we run the AGNT2 sidecar, the primary deployment primitive, with \texttt{\small agnt2/sidecar:latest} alongside it. We require zero \textit{framework-level} code change in the application container, since the sidecar owns the full on-chain interface; this claim holds directly for containers whose I/O is interceptable via HTTP or stdin/stdout. Containers using gRPC, WebSocket, or binary-protocol transports require a thin I/O adapter shim alongside the sidecar to translate between the transport and the sidecar's interception boundary, but the application logic and business code within those containers remain entirely unchanged.

Sequencer events for \texttt{\small callee == self.agent\_id} reach the sidecar through a Chain Listener, one of five modules, which keeps replay buffering and a polling fallback. We use the Identity Manager to publish registration, capability updates, heartbeats, and deregistration, leaving the registry as a typed service directory. The Payment Engine verifies incoming escrow, attaches outgoing payment, handles refunds, and manages Layer Top channels. Before invoking the local HTTP/gRPC service, the Interaction Router validates INVOKEs, fetches DA parameters, and maps schemas; it then stores the result and submits RESPOND. The Attestation Engine signs each RESPOND over interaction ID, input/output hashes, duration, timestamp, and image digest. We can re-execute deterministic containers against DA-stored inputs to verify outputs. That audit path is unavailable for LLM containers at temperature > 0 (\secref{subsec:architectural-limitations}).

\appref{app:sidecar-failure} covers deployment models, failure handling, the interaction lifecycle, CPU overhead, and Figure~\ref{fig:sidecar}'s module-boundary view, while leaving the application container unchanged. In particular, \appref{app:da-post-failure}--\appref{app:concurrent-open-channel} give the atomicity invariants, write-ahead-log (WAL) backed recovery path, and channel-id derivation for sidecar failure cases.

\subsection{Layer Top: P2P Fast Path}\label{subsec:layer-top}

Layer Top lets established pairs exchange thousands of signed sidecar-to-sidecar interactions without sequencer involvement [17]. Lifecycle: (1) both agents \texttt{\small OPEN\_CHANNEL} on Layer Core with bilateral deposits and TTL; (2) p2p \texttt{\small ChannelMessage} increments nonce and updates balances per interaction (full field specification in \appref{app:transaction-grammar}); (3) periodic \texttt{\small CHECKPOINT} compresses up to 1,000 interactions into one Layer Core transaction carrying the latest co-signed state; (4) cooperative or unilateral \texttt{\small CLOSE\_CHANNEL}, where unilateral close opens a 4-hour challenge window. Layer Core serializes concurrent \texttt{\small OPEN\_CHANNEL} races by block-sequence position: the second transaction is rejected with \texttt{\small CHANNEL\_EXISTS}, and the losing sidecar's deposits are credited to the canonical channel within one block. When a channel is unavailable, underfunded, unresponsive, or insufficient for a multi-party interaction, traffic escalates to Layer Core; \secref{subsec:layer-routing} summarizes that routing policy.

\subsection{Layer Core}\label{subsec:layer-core}

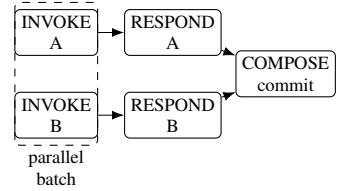
\begin{figure}[H]
\centering
\begin{tikzpicture}[
  font=\scriptsize,
  tx/.style={draw, rounded corners=2pt, align=center, inner sep=2.3pt,
             minimum width=1.1cm, minimum height=0.38cm},
  arrow/.style={-{Latex[length=1.6mm]}, line width=0.35pt}
]
\node[tx] (i1) at (0,0.55) {INVOKE\\A};
\node[tx] (i2) at (0,-0.55) {INVOKE\\B};
\node[tx] (r1) at (1.55,0.55) {RESPOND\\A};
\node[tx] (r2) at (1.55,-0.55) {RESPOND\\B};
\node[tx] (c) at (3.05,0) {COMPOSE\\commit};
\draw[arrow] (i1) -- (r1);
\draw[arrow] (i2) -- (r2);
\draw[arrow] (r1) -- (c);
\draw[arrow] (r2) -- (c);
\draw[dashed, rounded corners=2pt] (-0.55,-0.95) rectangle (0.55,0.95);
\node[align=center] at (0,-1.25) {parallel\\batch};
\end{tikzpicture}
\caption{Dependency-aware batching exposes parallelism while preserving INVOKE-before-RESPOND and COMPOSE-after-constituents ordering.}
\label{fig:batching}
\end{figure}

The Layer Core sequencer is the interaction rollup's central innovation; its pipeline is \texttt{\small Mempool → Dependency Analyzer → Batch Builder → Executor → Block Producer}. The \textbf{Dependency Analyzer} builds a DAG over pending transactions: a RESPOND depends on its INVOKE; a COMPOSE on its constituents; a CHECKPOINT on the channel; independent INVOKEs parallelize. Each transaction \textit{declares} predecessors at submission, so DAG construction is \textbf{O(n)} --- one edge per declared dependency --- versus O(n²) conflict detection used in Vegeta [8]; at 250K tx/block, O(n²) is infeasible while O(n) remains tractable. Batches preserve the INVOKE-before-RESPOND invariant and execute independent interactions in parallel (Figure~\ref{fig:batching}). The formal serialization-equivalence proof is follow-on work (\secref{subsubsec:formal-specification}).

The Layer Core sequencer, in Phase 1, is trusted-for-liveness and untrusted-for-correctness, much like Optimism's initial deployment. We require it to include transactions and publish pre-states to DA; its state-root submissions remain provisional until the Type 1 fraud-proof window closes. Correctness mostly depends on one condition: during the 1-hour challenge window, at least one honest party with DA access watches submitted roots, under the standard 1-of-N optimistic-rollup assumption. The sequencer posts a bond to Layer Root. If it equivocates, censors beyond TTL, or faces a successful fraud-proof challenge, that bond is slashed. In Phase 3, a rotating BFT committee over block headers replaces the single sequencer; we then move the assumption from 1-of-N monitoring to a BFT threshold.

AGNT2 exposes only typed interaction-trie transitions, not the EVM interface: INVOKE, RESPOND, COMPOSE, DISCOVER, REGISTER, TIMEOUT, channel ops, and DISPUTE. Before any trie mutation, we check format and signatures; payment and capability existence are checked there as well. INVOKE creates a pending record and locks escrow. RESPOND completes it, releases escrow, and updates reputation; COMPOSE remains an atomic batch. On TTL expiry, the executor emits TIMEOUT. Execution costs roughly 5--10 $\mu$s per transaction, versus 1--10 ms for a typical EVM transaction. With dependency-aware parallel scheduling, that 100$\times$--1000$\times$ reduction underwrites the 300K--500K TPS Layer Core design target on commodity hardware; no end-to-end Layer Core measurement yet exists (\secref{subsec:layer-core-execution}, C9). \appref{app:component-breakdown} gives the component breakdown; \appref{app:transaction-grammar} gives the transaction grammar, capability-weighted gas, and composition patterns.

The Layer Core state root is the Merkle root over the \textit{interaction trie}, which encodes agent identity, capabilities, reputation, economics, and interaction history as first-class protocol objects organized under per-agent, per-channel, per-interaction, per-dispute, and global namespaces; reputation counters derive deterministically on-chain without an aggregator. The full field catalogue and sizing assumptions are in \appref{app:interaction-trie}. The schema strictly supersets proposed ERC-8004 (early-stage; not yet an accepted EIP), preserving cross-chain interoperability. DISCOVER queries by capability, schema hash, price range, and minimum reputation, forming an autonomous on-chain capability market.

\textbf{Layer Core Ordering Guarantee.} Layer Core commits only blocks that satisfy a serialization-equivalence invariant: any topological ordering of the block's transaction DAG, executed by the deterministic transition function T, produces an identical final state. Declared dependency edges cover the common case (RESPOND names its INVOKE, COMPOSE names constituents), while a lightweight O(k) conservative-edge-addition pass catches undeclared same-object conflicts. Fraudulent \texttt{\small parent\_invoke\_id} values are checked during execution against PENDING trie pre-state; a mismatch renders the batch root invalid and triggers Type 1 fraud proof and sequencer slashing. Over-declaration is safe because widening the dependency set preserves serialization equivalence, while omission risks committing an invalid root. This two-pass construction --- O(n) declared edges + O(k) contention edges --- is what keeps ordering tractable at 250K tx/block. Formal proof of serialization equivalence is follow-on work (\secref{subsubsec:formal-specification}); the full invariant statement and edge-addition algorithm are in \appref{app:layer-core-ordering}.

\subsection{Layer Root: Settlement and Fraud Proofs}\label{subsec:layer-root}

Layer Root is AGNT2's settlement layer, a protocol component between Layer Core and the L1. It has two responsibilities: (1) batch Layer Core state roots to the L1 at adaptive frequency (10 min at <1K TPS; 2 min at 1K--500K TPS; 30 s at 500K+ TPS), carrying the covered block range, current/previous state roots, interaction counts, AGNT volume, and DA commitment; and (2) run the agent-typed re-execution VM for computational fraud proofs within the challenge window.

A root mismatch in a replayed batch proves computational fraud. For Type 1, we replay the challenged batch from supplied pre-state and transactions; if it diverges, we slash the sequencer and revert settlement. Its fraud-proof challenge window is 1 hour. Channel disputes instead use Type 2, with a 4-hour window, because stale states settle by nonce: the higher-nonce co-signed state wins, and same-nonce conflicts prove double-signing and slash both parties. Under partial synchrony, Type 1 timing is T\_DA (1--5 min sequential blob fetch) + T\_detect (<1 min re-execution) + T\_proof (<1 min) + T\_submit (<5 min L1 inclusion), about 8--12 min worst case. This leaves a 5--7$\times$ safety margin for DA outages and monitor restart. On a Layer Root-native L1 with BFT finality, we could shrink it to 10--15 min (full derivation in \appref{app:type1-window}).

On-chain transactions carry only a 32-byte \texttt{\small params\_hash} or \texttt{\small result\_hash}. We keep the full 2 KB message payloads on DA, where they must stay retrievable for the longest challenge window (4 hours). Before submitting a state-root, we require DA to be posted. Layer Root rejects any root without valid DA commitments, so withholding does not make T\_DA unbounded. The 500K TPS target counts INVOKE/RESPOND messages, which means DA write bandwidth of 500K $\times$ 2 KB $\approx$ 1 GB/sec, about two orders of magnitude above the current sustained envelopes for EigenDA (\textasciitilde{}10 MB/sec) and Celestia (\textasciitilde{}6 MB/sec). This \textasciitilde{}100$\times$ DA gap is the \secref{subsec:da-bandwidth-gap} operational ceiling addressed by the purpose-built agent-DA contribution in \secref{subsubsec:scalability-extensions}. Type 1/Type 2 reads stay sequential over batch pre-state and signed channel history, which avoids per-interaction random-access amplification. \appref{app:fraud-proof-derivations} gives the governance parameters and VM determinism across parallel batch schedulings. It also specifies DA read patterns and L1 integration modes.

\textbf{Fraud-proof scope by service type.} Type 1 proves \textit{state-machine compliance} (escrow sequencing, trie mutations, payment flow, INVOKE-before-RESPOND ordering) for all services. For \textit{deterministic} containers, DA-stored inputs further allow output re-execution. For \textit{nondeterministic} services (LLM inference at temperature > 0), only compliance is provable --- semantic output correctness is out of scope (\secref{subsec:architectural-limitations}, \secref{subsubsec:qualitative-adjudication}). \appref{app:l1-integration} gives the full per-type boundary.

\subsection{Layer Routing}\label{subsec:layer-routing}

An interaction that lacks an open channel with its target does not use Layer Top. We send an outgoing interaction there only when deposits are non-exhausted, the target is responsive, and the interaction is bilateral (not \texttt{\small COMPOSE}); otherwise we route it through Layer Core. \texttt{\small CHECKPOINT} and \texttt{\small SETTLE} batch upward, so most interaction volume stays on Layer Top off-sequencer. For first encounters, multi-party compositions, and verifiable records, Layer Core handles the traffic. Layer Root receives periodic settlements. Agents join AGNT2 on Layer Core. As channels open, traffic moves to Layer Top, mainly because of the \textasciitilde{}1000$\times$ per-interaction cost gap. Agents do not choose layers manually.

\begin{table*}[!t]
\centering
\scriptsize
\caption{Sequencer simulation results across workload scales. All results are simulation outputs using synthetic latency distributions; real agent execution latency variance is wider than the model assumes [simulation].}
\label{tab:sequencer}
\begin{adjustbox}{max width=\textwidth}
\begin{tabular}{@{}lrrrrrrr@{}}
\toprule
Workload & N & FIFO batches & DAG batches & FIFO tx/s & DAG tx/s & Speedup & DAG parallelism \\
\midrule
mixed-100 & 100 & 100 & 6 & 2.7 & 9.0 & 3.3$\times$ & 16.67$\times$ \\
mixed-1000 & 1,000 & 1,000 & 9 & 2.9 & 23.4 & 8.1$\times$ & 111$\times$ \\
mixed-10000 & 10,000 & 10,000 & 10 & 3.1 & 123.1 & 39.7$\times$ & 1,000$\times$ \\
real-crewai & 8 & 8 & 4 & --- & --- & 2.0$\times$ & 2.0$\times$ \\
\bottomrule
\end{tabular}
\end{adjustbox}
\end{table*}

\section{Security and Threat Model}\label{sec:security}

\subsection{Threat Model}\label{subsec:threat-model}

A Byzantine Layer Core sequencer can re-order, suppress, or front-run INVOKE/RESPOND transactions while ordering them into blocks under a known identity. It still cannot forge signatures, and fraudulent transitions that are state-root-consistent cannot pass undetected. Around that core, we model agents as callers or callees that do not trust each other; each agent holds a signing key and runs a container behind a sidecar. We use Layer Root validators to settle state roots to L1, inheriting the L1 honest-majority assumption. DA layer operators provide advertised liveness and retrievability under their service agreement. The underlying L1 chain is outside our scope; we assume it provides state-root storage and slashing enforcement, with eventual finality according to its consensus guarantees.

\subsection{Representative Threats (Adversary-Mechanism Summary)}\label{subsec:representative-threats}

A Byzantine sequencer front-running in T1 can still try the simple move: reorder pending INVOKEs. The load-bearing defenses constrain that move with an encrypted mempool run by the threshold-decryption committee (t-of-n, t = $\lceil$2n/3$\rceil$, n = 16--64), plus commit-reveal for price-sensitive capabilities. L1 force-inclusion remains the backstop [4, 5, 36]. Type 1 fraud proofs rule out state-root fraud in all configurations. We still assume at least $\lceil$2n/3$\rceil$ of n committee members are honest-for-decryption at each block interval. In T2, colluding agents manufacture reputation, but N synthetic-interaction colluders run into counterparty-diversity weighting, where diversity saturates and score growth is at most log-linear; a per-identity stake floor makes K manufactured identities cost K $\times$ stake\_min, and Zipf-violating traffic-pattern anomaly detection adds a statistical check. This is economic and statistical deterrence, not a cryptographic guarantee. A sufficiently capitalized adversary willing to incur stake costs is bounded but not excluded. For T3, a malicious sidecar may false-attest a DA result. At RESPOND ingestion, a protocol-level DA-commitment pre-check, together with a Layer-Root sample-retrieval check at settlement, prevents claims to a \texttt{\small result\_hash} without a retrievable DA blob. Downstream COMPOSE participants hash-check before committing. Here we assume DA-layer liveness within the bounded retrieval window (honest majority of DA operators) [14, 15]. \Appref{app:extended-threat-model} gives the full adversary-capability-mechanism-residual analyses for T1--T3.

\subsection{Threat Coverage}\label{subsec:threat-coverage}

Standard mitigations for service fraud, flooding, censorship, and Sybil attacks follow the rollup-L2 and service-mesh literature; \appref{app:extended-threat-model} gives the full adversary-capability-mechanism-residual analysis. We group mitigations into four families: cryptographic ordering defenses, economic/statistical deterrence, rate-limit quotas, and auditability mechanisms. \Appref{app:extended-threat-model} also maps nine safety properties to adversary model and evidence type.

\section{Evaluation}\label{sec:evaluation}

The claim-to-evidence map draws the line between claims we measured in the current prototype and claims we test through synthetic workload simulation, analytical capacity modeling, or leave design-only. The 300K-500K TPS Layer Core execution ceiling sits on the design side of that line. It is not a demonstrated result. For the remaining claims, we use prototype measurement, simulation, or closed-form analysis. Table~\ref{tab:da-summary} previews the DA-bounded deployment regimes discussed in \secref{subsec:da-bandwidth-gap}.

\begin{table*}[!t]
\centering
\footnotesize
\caption{DA regime summary. Canonical parameters and the first binding bottleneck at each throughput tier.}
\label{tab:da-summary}
\begin{tabularx}{\textwidth}{@{}YYY@{}}
\toprule
Parameter & Value & Source \\
\midrule
Canonical payload (one direction: INVOKE \textit{or} RESPOND) & 1 KB & \secref{subsec:da-bandwidth-gap}, \secref{subsec:escrow-gas} linear model \\
Write amplification (INVOKE + RESPOND both to DA) & 2$\times$ & \secref{subsec:layer-core} protocol \\
Effective DA write per completed interaction & 2 KB & derived \\
Layer Top off-sequencer bilateral offload & 65\% & \secref{subsec:layer-top}, \secref{subsec:escrow-gas} channel model \\
Effective DA demand (fraction of raw traffic) & 35\% & derived \\
Retention window (longest) & 4 hours & \secref{subsec:layer-root} Type 2 challenge \\
Near-term deployable backend & EigenDA (\textasciitilde{}10 MB/s) & \appref{app:component-breakdown} \\
DA read access pattern & sequential (batch pre-state, channel history) & \appref{app:da-read-patterns} \\
Near-term TPS regime & 10K--100K TPS & DA-bound \\
500K TPS design target requires & \textasciitilde{}1 GB/s & 100$\times$ above EigenDA \\
Path to target & purpose-built Agent-DA (\secref{subsubsec:scalability-extensions}) & design-only \\
\bottomrule
\end{tabularx}
\end{table*}

\setcounter{subsection}{-1}
\subsection{Evidence Overview}\label{subsec:evidence-overview}

Appendix H contains the complete C1--C10 claim-to-evidence map, with evidence type, strength, subsection, and upgrade path recorded for each claim. In this section, prototype measurement backs five claims (C5--C8, C10), simulation backs three (C1--C3), analysis backs one (C4), and one remains design-only (C9).

\subsection{Dependency Analyzer and Sequencer Parallelism}\label{subsec:dependency-analyzer}

We build a DAG over pending transactions by following \texttt{\small parent\_invoke\_id} edges; the algorithm is O(n) in pending transactions --- one edge per declared dependency versus O(n²) conflict detection in Vegeta [8]. AGNT2's central throughput claim is that dependency-aware DAG batching exposes substantially more parallelism than FIFO on realistic agent workloads. We simulate three synthetic workloads (mixed-100, mixed-1000, mixed-10000) and three real framework traces (AutoGen, CrewAI, LangGraph) under sequential FIFO and DAG-parallel batching [simulation]. The simulation clock sets each batch's wall-clock duration to the maximum per-transaction latency in that batch; transactions within a batch run in parallel. FIFO uses one transaction per batch, serializing all work independent of dependencies.

At N=100, DAG batching reduces 100 FIFO batches to 6 DAG batches, increasing throughput from 2.7 tx/s to 9.0 tx/s, a 3.3$\times$ speedup with 16.67$\times$ realized parallelism. Deep dependency chains bound the critical path, creating the gap between speedup and parallelism. At N=1,000, 1,000 FIFO batches compress to 9 DAG batches (23.4 tx/s, 8.1$\times$ over FIFO, 111$\times$ realized parallelism). At N=10,000, 10,000 FIFO batches collapse to 10 DAG batches (123.1 tx/s, 39.7$\times$ over FIFO, 1,000$\times$ realized parallelism). The plateau at roughly 10 DAG batches for N=1,000 and N=10,000 reflects the synthetic workload's average dependency depth, which limits critical-path length regardless of total transaction count. Table~\ref{tab:sequencer} summarizes the full workload-scale results.

The real framework traces show the parallelism current agent programs actually expose. AutoGen (3 agents, 23 interactions, chain depth 23) is fully sequential: each step depends on the previous output, so DAG batching and FIFO yield equivalent throughput. CrewAI (5 agents, 8 interactions) realizes 2.0$\times$ parallelism from fan-out tasks that form two independent batches. LangGraph (6 agents, 6 interactions) reaches 1.2$\times$ parallelism from one parallel retrieval pair. Thus, 1,000$\times$ simulation gains require workloads deliberately structured for independent parallel work; production agent programs currently exploit only a narrow slice of available concurrency. The simulation also confirms that the Batch Builder correctly emits dependency-ordered blocks (C3): a 10,000-transaction workload produces 10 blocks under DAG scheduling rather than 10,000 FIFO blocks. Upgrading this claim requires operating a distributed sequencer cluster under fault injection --- leader failover mid-batch, dropped block proposals, and network partition. The simulation uses synthetic latency distributions calibrated to observed p95 values; wider real-world variance, such as AutoGen's 44,759 ms p95, will reduce realized speedup relative to the synthetic model. \Appref{app:layer-core-ordering} gives the full ordering invariant and two-pass edge-construction rule that the simulation assumes.

\subsection{Data Availability Bandwidth Gap}\label{subsec:da-bandwidth-gap}

The DA constraint on AGNT2 throughput is characterized analytically using a linear calldata model verified against all six measured gas data points from \secref{subsec:escrow-gas} [analytical]. Calldata per workflow scales as: \texttt{\small createWorkflow = 164 + 64×N bytes}, \texttt{\small settle = 100 + 256×N bytes}, total \texttt{\small = 264 + 320×N bytes}. At N=10 agents, this yields 3,464 bytes (3.4 KB) per workflow. EIP-4844 blob encoding substantially reduces the cost of posting this data to L1: at N=10, L1 calldata at 10 gwei costs \$1.108 per workflow, while blob posting at 1 gwei costs \$0.007 --- a 156.5$\times$ reduction that makes DA economically tractable at moderate throughput. The full field-width derivation and per-payload saturation ceilings across DA backends are in \Appref{app:throughput-analysis} (Table~\ref{tab:da-ceilings}); deployment regimes by TPS under the canonical 1 KB/interaction, 2$\times$ write-amplification model are in \appref{app:throughput-analysis} (Table~\ref{tab:da-regimes}).

Celestia (6 MB/s sustained) saturates at 6,144 TPS and EigenDA (10 MB/s) at 10,240 TPS at 1 kB per workflow; a purpose-built Agent-DA layer (200 GB/s design target) would support 204.8M TPS --- five orders of magnitude above current deployable infrastructure. With Layer Top carrying 65\% of bilateral interactions off-chain, effective DA demand is 35\% of raw. The practical consequence is the \textasciitilde{}100$\times$ gap that constrains near-term deployment: 500K TPS $\times$ 2 KB per workflow $\approx$ 1 GB/sec DA throughput, whereas EigenDA provides \textasciitilde{}10 MB/sec sustained. Near-term deployments are DA-tractable at 10K TPS under existing EigenDA/Celestia; the 50K--100K TPS tier requires near-future DA capacity (Table~\ref{tab:da-regimes}), and reaching the Layer Core ceiling requires the purpose-built DA layer in \secref{subsubsec:scalability-extensions}. Table~\ref{tab:da-summary} records the canonical parameters and per-tier bottlenecks.

One access-pattern concern deserves a direct answer: an obvious reviewer worry is that fraud-proof verification would hit DA with random-access reads by interaction hash, compounding the bandwidth gap against blob-oriented backends. It does not. Type 1 computational fraud proofs re-execute the full disputed batch against its posted pre-state, and Type 2 channel disputes replay signed channel history from the CHECKPOINT co-signed state --- both are sequential reads over contiguous data, matching how Celestia and EigenDA are already optimized. The access-pattern derivation is in \appref{app:da-read-patterns}.

\subsection{Zero-Code Sidecar Integration}\label{subsec:zero-code-sidecar}

The sidecar wraps any agent process by intercepting its stdout/stdin streams; no modification to the agent framework or application code is required [measured]. Integration was validated against three unmodified frameworks: AutoGen (3 agents, 23 interactions, avg request payload 1,009 bytes, p95 latency 44,759 ms, chain depth 23), CrewAI (5 agents, 8 interactions, avg response payload 3,904 bytes, p95 latency 46,156 ms, chain depth 4, parallelism 2.0$\times$), and LangGraph (6 agents, 6 interactions, avg response payload 3,080 bytes, p95 latency 3,407 ms, chain depth 5, parallelism 1.2$\times$). All three completed their respective workloads without framework modification. \Appref{app:sidecar-architecture} and \appref{app:interaction-lifecycle} give the sidecar module view and per-step interaction lifecycle behind this wrapping model. Upgrading this claim requires testing against a wider set of frameworks --- particularly those with non-standard I/O patterns or multiplexed streams --- and reporting compatibility failures, not just successes.

\subsection{Channel Open/Close Latency}\label{subsec:channel-latency}

The prototype measures state channel lifecycle gas costs and wall-clock latency on a local Anvil instance [measured]. \texttt{\small openChannel} consumes 179,368 gas at 94 ms wall-clock; \texttt{\small cooperativeClose} consumes 134,859 gas at 85 ms; \texttt{\small initiateUnilateralClose} consumes 109,404 gas at 84 ms. All three operations fall below the 100 ms target on loopback Anvil. These figures exclude network RTT: WAN deployments add at least one round-trip per on-chain confirmation, pushing observed latency above 100 ms in most cloud regions. A definitive characterization requires measurement under WAN conditions with concurrent channel churn.

\subsection{Escrow Gas Cost}\label{subsec:escrow-gas}

The escrow contract is the most extensively measured component of the prototype. Gas consumption for both \texttt{\small createWorkflow} and \texttt{\small settle} was measured across six values of N (the number of participating agents) on Anvil, with all transaction hashes recorded [measured]. The results confirm near-perfect linear scaling in N for both operations; Table~\ref{tab:escrow} records all measurements.

\begin{table*}[!t]
\centering
\footnotesize
\caption{Escrow gas measurements across workflow sizes, with N = agent count; we measured all figures on Anvil localhost [measured].}
\label{tab:escrow}
\begin{adjustbox}{max width=\textwidth}
\begin{tabular}{@{}rrrrr@{}}
\toprule
N & createWorkflow gas & settle gas & total gas & Cost at 1 gwei, ETH=\$3,000 \\
\midrule
1 & 243,736 & 97,516 & 341,252 & \$1.02 \\
3 & 318,122 & 178,043 & 496,165 & \$1.49 \\
5 & 409,645 & 258,547 & 668,192 & \$2.00 \\
10 & 638,379 & 459,865 & 1,098,244 & \$3.29 \\
15 & 867,101 & 661,171 & 1,528,272 & \$4.58 \\
20 & 1,095,824 & 862,496 & 1,958,320 & \$5.87 \\
\bottomrule
\end{tabular}
\end{adjustbox}
\end{table*}

OLS regression on the six data points gives \texttt{\small createWorkflow = 188,239 + 45,248×N}, \texttt{\small settle = 57,249 + 40,262×N}, and \texttt{\small total = 245,488 + 85,510×N}. In all three fits, the R² is approximately 1.00. That result shows the Solidity implementation is cost-linear in N, with no detectable super-linear term. The contract enforces \texttt{\small MAX\_STEPS = 20}, capping total escrow gas at 1,958,320 (approximately \$5.87 at 1 gwei and ETH=\$3,000). On an L2 operating at 0.1 gwei (an aggressive but achievable fee floor), a N=10 workflow costs \$0.33. The per-agent marginal gas cost is 85,510 gas per additional participant, split roughly equally between the \texttt{\small createWorkflow} (45,248 gas) and \texttt{\small settle} (40,262 gas) contributions. \Appref{app:calldata-derivation} gives the calldata field-width derivation that links these gas measurements to the byte model used in \secref{subsec:da-bandwidth-gap}.

\subsection{COMPOSE Atomic Settlement}\label{subsec:compose-settlement}

All 16/16 Foundry tests pass in the current run. At the contract level, we verify atomic settlement and escrow rollback [measured]. In the \texttt{\small settle()} function, we read all N agent signatures from calldata; when any signature is missing or invalid, the call reverts. Until every signature arrives, escrow stays locked. If they do not arrive, the \texttt{\small timeoutSec} challenge window expires, and the initiating party may reclaim funds. The Foundry test suite exercises this path in the T4--T9 test cases: invalid signature reverts and window expiry with time-advance are both covered there. We also check nonce ordering enforcement. \Appref{app:atomic-composition} gives the formal COMPOSE semantics that these tests exercise at the contract level. Rollback under Byzantine counterparty behavior and sequencer-induced fault injection is not yet tested; those scenarios remain the upgrade path for this claim.

\subsection{Layer Core Execution Ceiling}\label{subsec:layer-core-execution}

No Layer Core prototype exists [design-only], so the 300K--500K TPS Layer Core execution ceiling remains a design target from first-principles decomposition. The single-threaded ceiling of 100K--200K TPS follows from a 5--10 $\mu$s per-transaction budget, dominated by hardware-accelerated Ed25519 signature verification and trie I/O. With software-only Ed25519, the cost expands to \textasciitilde{}50 $\mu$s and the ceiling falls to \textasciitilde{}20K TPS. If four worker threads run on disjoint trie-node batches, as the Dependency Analyzer permits, parallel execution raises the theoretical ceiling to 400K--800K TPS; the 500K target then sits inside that parallel window. We count typed-VM execution only. DA write bandwidth, inter-node replication, and networking overhead are excluded here and modeled separately (\secref{subsec:da-bandwidth-gap}, \appref{app:component-breakdown}). \Appref{app:component-breakdown} gives the per-component execution-budget breakdown behind this estimate. Evidence for the claim requires building and benchmarking an end-to-end Layer Core prototype. The claim would be falsified if, on such a prototype running a realistic AGNT2 agent workload --- with hardware-accelerated Ed25519, trie I/O, multi-worker parallel scheduling, and network overhead all included --- sustained throughput fell materially below 100K TPS on commodity server hardware; that would indicate that the 5--10 $\mu$s per-transaction budget does not hold in practice and that the 500K ceiling is unreachable with the proposed architecture.

\subsection{Sidecar Crash Recovery}\label{subsec:sidecar-crash-recovery}

The prototype implements unilateral channel close with a 24-hour challenge window --- a conservative Anvil testbed placeholder, not the protocol value; the production Layer Top / Type 2 window is \textbf{4 hours} (\secref{subsec:layer-top}, \secref{subsec:layer-root}, \appref{app:type1-window}). \texttt{\small finalizeUnilateralClose} is measured at 68,722 gas on Anvil, and the EVM time-advance suite verifies the lockout enforcement [measured]. \Appref{app:da-post-failure}--\appref{app:concurrent-open-channel} gives the deferred WAL recovery and failure-path state machines behind the crash-recovery upgrade path. SQLite-backed process-death recovery --- allowing a sidecar to reconstruct pending escrow commitments after an unclean shutdown --- is follow-on work; long-duration crash and partition tests are the upgrade path.

Across R1--R7, the evidence pattern is now clear. R2 (rich payload) and R5 (micropayment channels) are supported by the DA-plus-channel design and demonstrated in the prototype (\secref{subsec:zero-code-sidecar}--\secref{subsec:escrow-gas}). R1 (sub-second latency), R3 (dependency-aware ordering), and R6 (COMPOSE atomicity) are supported at the component level: Layer Top latency is measured on Anvil (\secref{subsec:channel-latency}), the Dependency Analyzer is validated in simulation (\secref{subsec:dependency-analyzer}), and COMPOSE atomic rollback is verified by Foundry tests (\secref{subsec:compose-settlement}). R4 (interaction trie) and R7 (computational fraud proofs) remain design-only at full scope; the prototype validates only the escrow component of R7 (\secref{subsec:escrow-gas}--\secref{subsec:compose-settlement}), while the fraud-proof VM (\appref{app:fraud-proof-bisection}) and trie schema (\appref{app:interaction-trie}) are not yet implemented. \Appref{app:requirements-satisfaction} records the full R1--R7 mapping.

\subsection{Comparative Analysis and Residual Security Gaps}\label{subsec:comparative-analysis}

Among evaluated systems, AGNT2 is the only one \textit{designed} to satisfy all nine evaluation dimensions (\appref{app:appendix-comparative-analysis}); the comparison reflects architectural design targets, not end-to-end measurements. Kubernetes+Istio handles frequency and payload but fails on identity, payment, composition, and cross-principal trust. Existing L2s cover settlement and fraud proofs but fail all agent-specific dimensions --- their challenge windows re-execute EVM over 7 days vs. AGNT2's 1-hour typed-VM window (\appref{app:fraud-proof-bisection}). Fetch.ai and Autonolas satisfy subsets but leave interactions off-chain. Per-threat coverage is detailed in \appref{app:extended-threat-model}; the remaining medium-or-lower-coverage threats --- service fraud, collusion, data poisoning, agent MEV --- all exploit social or economic patterns that remain partly inside protocol rules, while LLM-class fraud still relies on reputational deterrence (\secref{subsubsec:qualitative-adjudication}).

\section{Implementation Considerations}\label{sec:implementation}

AGNT2 has a partial working prototype (\secref{subsec:evidence-overview}) that validates five focused claims. This section identifies reusable and net-new components for a production-grade implementation, explains the main build-versus-fork choice, and records prototype scope, residual engineering risks, and what remains design-only.

\subsection{Component Decomposition and Reuse}\label{subsec:component-decomposition}

AGNT2 decomposes into seven implementation components [12, 13, 14, 15, 17, 32, 33, 34]. Two reduce mostly to configuration: the EVM-compatible L1 anchor and DA SDK integration. Three are partial reuses: Layer Root settlement can fork OP Stack/Orbit contracts while replacing EVM re-execution with an agent-typed fraud VM (\appref{app:fraud-proof-derivations}); the interaction trie adapts MPT/Verkle-style authenticated state to agent, channel, and interaction nodes (\appref{app:interaction-trie}); and Layer Top adapts Lightning-style bilateral channels from payments to service invocations (\appref{app:transaction-grammar}, \appref{app:concurrent-open-channel}). Two are net-new and constitute the paper's main implementation contribution: the Layer Core typed VM with dependency-aware batching (\appref{app:transaction-grammar}, \appref{app:component-breakdown}), and the chain-aware sidecar (\appref{app:sidecar-failure}).

\subsection{Build-versus-Fork Decisions}\label{subsec:build-vs-fork}

Forking OP Stack or Arbitrum Orbit inherits mature L1 settlement, fraud-proof messaging, and block-production machinery, but it also inherits EVM execution and single-threaded block processing. That path is plausible for an MVP below 100K TPS, but not for the 300K--500K \secref{subsec:layer-core-execution} design envelope. The mechanism-level gap is twofold: (i) a disputed AGNT2 step forces the EVM verifier to re-execute trie traversal + typed-record decoding + state-machine update in Solidity (witness cost scales with trie ops, not one native VM step; \appref{app:fraud-proof-bisection} table); (ii) throughput --- precompile invocation costs \textasciitilde{}700--3,000 gas (\textasciitilde{}0.07--0.3 ms at 10M gas/s), per-transaction signature recovery/nonce/trie updates cost 21K--50K gas (\textasciitilde{}2--5 ms), and intra-block execution is serial, jointly capping EVM-hosted AGNT2 at \textasciitilde{}1K--10K TPS. The native path instead budgets 5--10 $\mu$s per operation (\appref{app:component-breakdown}) and a one-step terminal check. Native typed transactions (\secref{subsec:layer-core}) are therefore required; EVM precompiles are not a viable path.

\subsection{Prototype Scope, Residual Risks, and What Remains Design-Only}\label{subsec:prototype-scope}

Prototype coverage is intentionally narrow. Anvil runs and Foundry tests support five claims (C5--C8, C10): zero-code sidecar integration (\secref{subsec:zero-code-sidecar}), channel lifecycle measurements (\secref{subsec:channel-latency}, \secref{subsec:sidecar-crash-recovery}), escrow gas linearity (\secref{subsec:escrow-gas}), and COMPOSE atomic rollback (\secref{subsec:compose-settlement}). Simulation supports C1--C3, C4 remains analytical, and C9 remains design-only. \Appref{app:claim-to-evidence} records the full claim-to-evidence map and upgrade paths.

Three implementation risks remain after the \secref{sec:architecture}--\secref{sec:evaluation} design choices:

\begin{itemize}
  \item \textbf{Sequencer skew.} \secref{subsec:dependency-analyzer} bounds heavy-Zipf throughput at 200K--300K TPS with agent sharding; \appref{app:hot-agent-conflicts} gives the hot-agent conflict and cache-footprint analysis. Dynamic rebalancing during popularity shifts remains production engineering.
  \item \textbf{Data availability.} Current Celestia/EigenDA envelopes bound near-term deployments to 10K--100K TPS; the full 500K TPS target depends on the agent-DA layer in \secref{subsubsec:scalability-extensions}. \Appref{app:throughput-analysis} gives the saturation derivations and throughput-regime classification.
  \item \textbf{Sidecar scale.} The \secref{subsec:sidecar-pattern} budget is <1 ms CPU per interaction up to \textasciitilde{}10K inbound INVOKEs/sec/host; \appref{app:cpu-overhead} gives the CPU-overhead breakdown. Larger fleets need connection multiplexing and shared chain-state caches using established Envoy/Linkerd patterns.
\end{itemize}

The Layer Core typed VM remains design-only (\appref{app:fraud-proof-bisection}), as do the distributed sequencer and Layer Root fraud VM. We treat the \secref{subsec:dependency-analyzer} throughput envelope and \secref{subsec:da-bandwidth-gap} DA regime analysis as analytical projections, not end-to-end stack measurements. The failure-boundary state machines specified in \appref{app:da-post-failure}--\appref{app:concurrent-open-channel} --- mid-channel sidecar crash recovery, concurrent \texttt{\small OPEN\_CHANNEL} resolution, DA-post-failure recovery, and malformed-LLM-output handling --- are design-specified but have not been adversarially stress-tested; long-duration crash, partition, and race-condition campaigns are follow-on work. We likewise defer formal ordering proofs (\secref{subsubsec:formal-specification}) and cryptoeconomics (\secref{subsubsec:cryptoeconomic-design}--\secref{subsubsec:qualitative-adjudication}). \Appref{app:throughput-analysis} contains the broader throughput-regime analysis and DA saturation derivations.

\section{Discussion and Future Work}\label{sec:discussion}

\subsection{Architectural Limitations}\label{subsec:architectural-limitations}

Layer Top's sub-100 ms path still cannot outrun DA-posting latency: 10--50 ms sits on the finality critical path, before Layer Core adds its 500 ms block interval. We therefore do not aim AGNT2 at sub-millisecond deadlines such as HFT, real-time control, or hardware-in-the-loop. In the base configuration, Layer Core relies on a known sequencer that cannot produce fraudulent state roots (Type 1 detects them), but liveness depends on its availability; established channels fall back to Layer Top and L1 force-inclusion provides a mitigation. Espresso, Astria, and shared sequencers could fit; we leave those integrations unspecified. Sybil resistance is economic rather than cryptographic: per-identity stake and interaction-history cost make Sybils expensive, not impossible; reputation diversity-weighting (the T2 discussion in \secref{subsec:representative-threats}, \appref{app:reputation-trie}, \appref{app:threat-collusion}) and anomaly detection are the remaining deterrents. We reject permissioned consortium membership (Fabric, Cosmos zones) in \secref{subsec:why-onchain} because it does not fit the open agent economy. LLM non-determinism limits off-chain auditability: for containers at temperature > 0, auditors rely on lower-temperature re-runs and verification-agent corroboration rather than byte-identical reproduction (\appref{app:llm-nondeterminism}, \appref{app:l1-integration}).

\subsection{Design Tradeoffs}\label{subsec:design-tradeoffs}

\textbf{Optimistic vs. ZK fraud proofs.} AGNT2 uses optimistic Type 1 over a 1-hour window (\secref{subsec:layer-root}, \appref{app:type1-window}). ZK finalizes batches immediately but adds per-tx proving cost: viable at 10K-TPS DeFi scale, not 500K TPS absent specialized hardware. It remains consistent for lower-TPS deployments. \textbf{Sidecar vs. native chain integration.} Sidecars trade per-host overhead for zero container modification, suiting heterogeneous multi-team agents; homogeneous single-team fleets may choose native integration. \textbf{One-hour Type 1 window vs. shorter.} Shorter windows reduce lockup but tighten safety; Layer Root-native BFT finality could reach 10--15 min. One hour fits generic EVM L1 today.

\subsection{External Assumptions}\label{subsec:external-assumptions}

\secref{subsec:da-bandwidth-gap}'s DA-layer trajectory works only while purpose-built agent-DA runs at full capacity (\secref{subsubsec:scalability-extensions}); if Celestia/EigenDA plateau, deployments cap at 10K--100K TPS. Even after DA matches, 500K TPS remains the execution-layer ceiling. Because the Zipf model assumes microservice-RPC-like traffic, we still must rederive the parallel-execution analysis for heavier-tailed or more uniform traffic. \Appref{app:throughput-analysis} gives the current analytical regime split; \secref{subsubsec:empirical-workload} gives the empirical follow-on.

\subsection{Deferred Research Agendas}\label{subsec:deferred-research}

\subsubsection{Validation and Formalization}\label{subsubsec:empirical-workload}\label{subsubsec:formal-specification} We still need to validate \secref{subsec:dependency-analyzer} throughput parameters against RPC, DeFi, AutoGen, CrewAI, LangGraph, and Optimism/Arbitrum traces [12, 13, 24, 35], replacing the Zipf-model assumption with empirical workload distributions. We also still need formal specifications for O(n) DAG serialization equivalence, VM semantics, trie-pruning adversarial bounds, and cross-L2 capability ontology. Together, these are the empirical and proof-theoretic follow-ons to \appref{app:throughput-analysis}, \secref{subsec:layer-core}, and \appref{app:layer-core-ordering}.

\subsubsection{Incentives, Governance, and Adjudication}\label{subsubsec:cryptoeconomic-design}\label{subsubsec:qualitative-adjudication} Capability-weighted gas, channel economics, Sybil bounds, dispute bonding, and token mechanics remain open protocol-design questions [20, 21, 22, 29]. Beyond that, semantic-fraud adjudication, LLM nondeterminism handling, and L1$\rightarrow$Layer Core governance propagation remain deferred. The present paper only bounds these issues at a high level: it proves computational correctness, not semantic output quality, and leaves the incentive and governance layer to future work.

\subsubsection{Security and Scalability Completion}\label{subsubsec:security-completion}\label{subsubsec:scalability-extensions} Agent MEV characterization, game-theoretic collusion bounds, and censorship-resistance liveness proofs remain incomplete, and \Appref{app:extended-threat-model} states the current threat model rather than the final one. On the systems side, agent-optimized DA with regional sharding and retention [14, 15], Phase 1 stack choice, L4 channels, cross-L2 composability, and the Layer Root-native L1 [12, 13, 17] are the main scale-out agenda. This is the path beyond the current DA-bound deployment envelope.

\section{Conclusion}\label{sec:conclusion}

Today's blockchain infrastructure still expects a human at the transaction boundary. Autonomous agents do not fit that model. With machine-speed interactions and structured payloads, token transfer becomes a poor center of gravity for composable multi-party invocations, so we use service invocation as the execution layer. AGNT2 proposes an agent-native Layer 2 system architecture built around that shift: Layer Top for fast bilateral channels, Layer Core for typed dependency-aware execution, and Layer Root for settlement and computational fraud proofs, with zero-code sidecar deployment and an interaction trie as first-class state.

The paper's contribution is therefore not only a new stack decomposition, but a narrower systems claim: modern wallets and delegation mechanisms increasingly solve authorization, while agent-native execution semantics remain missing from today's general-purpose chains. Our prototype evidence covers five core protocol claims (\secref{subsec:evidence-overview}, \secref{subsec:prototype-scope}), while the highest-throughput Layer Core ceiling and the full fraud-proof path remain design targets rather than demonstrated results. In the near term, data availability remains the main deployment bottleneck. Even so, we argue that if autonomous agents become durable economic actors, then service invocation, composition, escrow, and portable interaction history should be treated as protocol-native concerns rather than left entirely to application-layer convention.

\section*{References}

\begin{enumerate}
  \item Z. Ye, U. Misra, J. Cheng, W. Zhou, and D. Song, \textit{Specular: Towards Secure, Trust-minimized Optimistic Blockchain Execution}, IEEE S\&P, 2024.

  \item Z. Sun, Z. Li, X. Peng, X. Luo, M. Jiang, H. Zhou, and Y. Zhang, \textit{DoubleUp Roll: Double-spending in Arbitrum by Rolling It Back}, ACM CCS, 2024.

  \item C. F. Torres, A. Mamuti, B. Weintraub, C. Nita-Rotaru, and S. Shinde, \textit{Rolling in the Shadows: Analyzing the Extraction of MEV Across Layer-2 Rollups}, ACM CCS, 2024.

  \item P. Daian, S. Goldfeder, T. Kell, Y. Li, X. Zhao, I. Bentov, L. Breidenbach, and A. Juels, \textit{Flash Boys 2.0: Frontrunning in Decentralized Exchanges, Miner Extractable Value, and Consensus Instability}, IEEE S\&P, 2020.

  \item M. Kelkar, S. Deb, S. Long, A. Juels, and S. Kannan, \textit{Themis: Fast, Strong Order-Fairness in Byzantine Consensus}, ACM CCS, 2023.

  \item G. Danezis, L. Kokoris-Kogias, A. Sonnino, and A. Spiegelman, \textit{Narwhal and Tusk: A DAG-Based Mempool and Efficient BFT Consensus}, EuroSys, 2022.

  \item S. Mu, Y. Cui, Y. Zhang, W. Lloyd, and J. Li, \textit{Extracting More Concurrency from Distributed Transactions}, OSDI, 2014.

  \item T. Xu, Y. Zhong, Y. Zhang, R. Xiong, J. Zhang, G. Xue, and S. Liu, \textit{Vegeta: Enabling Parallel Smart Contract Execution in Leaderless Blockchains}, NSDI, 2025.

  \item K. Wüst, S. Matetic, S. Egli, K. Kostiainen, and S. Capkun, \textit{ACE: Asynchronous and Concurrent Execution of Complex Smart Contracts}, ACM CCS, 2020.

  \item J. Wang and H. Wang, \textit{Monoxide: Scale Out Blockchains with Asynchronous Consensus Zones}, NSDI, 2019.

  \item G. Ramseyer, A. Goel, and D. Mazieres, \textit{SPEEDEX: A Scalable, Parallelizable, and Economically Efficient Decentralized Exchange}, NSDI, 2023.

  \item Optimism Foundation, \textit{OP Stack Specification}, https://specs.optimism.io, 2022-2026.

  \item Offchain Labs, \textit{Arbitrum Nitro: A Second-Generation Optimistic Rollup}, https://docs.arbitrum.io/nitro-whitepaper.pdf, 2022.

  \item M. Al-Bassam, \textit{LazyLedger: A Distributed Data Availability Ledger With Client-Side Smart Contracts}, arXiv:1905.09274, 2019.

  \item Eigen Labs, \textit{EigenDA: The Hyperscale Verifiable Data Availability Layer}, https://docs.eigencloud.xyz/eigenda, 2023-2026.

  \item V. Buterin et al., \textit{EIP-4844: Shard Blob Transactions}, Ethereum Improvement Proposals, 2022.

  \item J. Poon and T. Dryja, \textit{The Bitcoin Lightning Network: Scalable Off-Chain Instant Payments}, 2016.

  \item Anthropic, \textit{Model Context Protocol Specification}, https://modelcontextprotocol.io, 2024-2026.

  \item Google and the Linux Foundation, \textit{Agent-to-Agent Protocol Specification}, https://a2a-protocol.org, 2025-2026.

  \item Valory AG, \textit{Autonolas Whitepaper v1.0}, 2023.

  \item M. Wooldridge et al., \textit{Fetch.ai: An Architecture for Modern Multi-Agent Systems}, arXiv:2510.18699, 2025.

  \item S. Zhou, K. Wang, and H. Yin, \textit{Dstack: A Zero Trust Framework for Confidential Containers}, arXiv:2509.11555, 2025.

  \item D. M. Rothschild, M. Mobius, J. M. Hofman, E. W. Dillon, D. G. Goldstein, N. Immorlica, S. Jaffe, B. Lucier, A. Slivkins, and M. Vogel, \textit{The Agentic Economy}, arXiv:2505.15799, 2025.

  \item M. Stein, \textit{How Are AI Agents Used? Evidence from 177,000 MCP Tools}, arXiv:2603.23802, 2026.

  \item \textit{Agent-OSI: A Six-Layer Reference Stack for Decentralized Agent Networking}, arXiv:2602.13795, 2026.

  \item \textit{UltraHorizon: Long-Horizon Agent Reasoning Benchmark}, arXiv:2509.21766, 2025.

  \item E. Androulaki et al., \textit{Hyperledger Fabric: A Distributed Operating System for Permissioned Blockchains}, EuroSys, 2018.

  \item Cosmos SDK, \textit{Cosmos SDK Documentation}, Online documentation, 2026.

  \item Ritual Network, \textit{Ritual Documentation}, Online documentation, 2026.

  \item Foundation for Intelligent Physical Agents, \textit{FIPA ACL Message Structure Specification}, 2002.

  \item T. Finin, R. Fritzson, D. McKay, and R. McEntire, \textit{KQML as an Agent Communication Language}, CIKM, 1994.

  \item Kubernetes, \textit{Kubernetes Documentation}, Online documentation, 2026.

  \item Istio, \textit{Istio Documentation}, Online documentation, 2026.

  \item Envoy Proxy, \textit{Envoy Documentation}, Online documentation, 2026.

  \item P. Moritz, R. Nishihara, S. Wang, A. Tumanov, R. Liaw, E. Liang, M. Elibol, Z. Yang, W. Paul, M. I. Jordan, and I. Stoica, \textit{Ray: A Distributed Framework for Emerging AI Applications}, OSDI, 2018.

  \item J. Bebel and D. Ojha, \textit{Ferveo: Threshold Decryption for Mempool Privacy in BFT Networks}, Cryptology ePrint Archive, Paper 2022/898, 2022.

  \item V. Buterin, D. C. A. Johnson, Y. Weiss, S. Bains, and others, \textit{EIP-4337: Account Abstraction Using Alt Mempool}, Ethereum Improvement Proposals, 2021-2026.

  \item Coinbase Developer Platform, \textit{Agentic Wallet Documentation}, https://docs.cdp.coinbase.com/agentic-wallet/welcome, 2026.

  \item M. De Rossi, D. Crapis, J. Ellis, and E. Reppel, \textit{EIP-8004: Trustless Agents}, Ethereum Improvement Proposals, 2025-2026.

  \item Coinbase Developer Platform, \textit{x402 Documentation}, https://docs.cdp.coinbase.com/x402/welcome, 2025-2026.

\end{enumerate}
The appendices hold the full technical details for reviewers who want them: protocol grammar, failure handling, derivations, schemas, extended related work, and evidence mapping.

\clearpage
\appendix
\titleformat{\section}{\normalsize\bfseries}{Appendix \thesection:}{0.6em}{}

\section{Transaction Grammar and Opcode Semantics}\label{app:transaction-grammar}

Gas is where the structure gets dense. capability-weighted Full Layer Core transaction grammar below. Atomic-composition semantics from there tie into the four patterns from \secref{subsec:layer-core}.

\subsection{Transaction Type Specification}\label{app:transaction-type-spec}

\begin{figure*}[t]
\begin{Verbatim}[fontsize=\scriptsize,baselinestretch=0.88,frame=single,framesep=3pt]
// Agent lifecycle
REGISTER { agent_id, name, capabilities[], stake, metadata }
DEREGISTER { agent_id }  // triggers unstake after cooldown
UPDATE_CAPABILITIES { agent_id, capabilities[] }

// Core interaction
INVOKE {
  id, caller, callee, capability, params_hash,
  payment, ttl, session_id?, composition_id?, da_pointer, nonce, sig
}
RESPOND {
  interaction_id, responder, result_hash, attestation,
  execution_ms, da_pointer, sig
}
TIMEOUT { interaction_id }  // auto-generated by sequencer

// Composition
COMPOSE {
  id, initiator, interactions: INVOKE[], dependency_graph,
  total_payment, atomic: bool
}

// Channels
OPEN_CHANNEL { channel_id, agent_a, agent_b, deposit_a, deposit_b, ttl, max_interactions }
CHECKPOINT { channel_id, nonce, balance_a, balance_b, interaction_count, sig_a, sig_b }
CLOSE_CHANNEL { channel_id, final_nonce, final_balance_a, final_balance_b, sig_a, sig_b }

// Disputes
DISPUTE { target, disputant, dispute_type: COMPUTATIONAL|QUALITATIVE|STALE_STATE, evidence_hash, da_pointer }

// Discovery
DISCOVER { capability_filter, schema_hash?, price_range?, min_reputation? }
\end{Verbatim}
\end{figure*}

By CHECKPOINT time, Layer Top ChannelMessage is up. It's the off-chain p2p format that carries the fields needed to reconstruct interaction record on-chain:

\begin{Verbatim}[fontsize=\footnotesize,baselinestretch=0.88]
ChannelMessage {
  channel_id:     bytes32
  nonce:          uint64          // monotonically increasing
  msg_type:       INVOKE | RESPOND
  capability:     string
  params_hash:    bytes32         // hash of full payload
  result_hash:    bytes32         // hash of response (RESPOND only)
  payment_delta:  int256          // balance change
  cumulative_a:   uint256         // A's running balance
  cumulative_b:   uint256         // B's running balance
  timestamp:      uint64
  sig_sender:     bytes
}
\end{Verbatim}

\subsection{Executor State Transitions}\label{app:executor-state-transitions}

For each transaction type, our executor validates format and signatures first. Only after we check payment sufficiency and capability existence do we apply the corresponding trie update.

INVOKE opens the pending interaction record and locks the escrowed payment. It then writes the TTL deadline. After the call finishes, RESPOND completes the same record; the escrow is paid to the callee, and the callee's reputation counters are updated. For registration, we let REGISTER add or update an agent node with identity and capabilities, with economics kept in the same entry. The awkward part is composition. COMPOSE runs it as one atomic batch, so any failed constituent INVOKE or RESPOND reverts the whole composition. When a TTL expires, the sequencer auto-generates TIMEOUT and refunds escrow to the caller. It also penalizes callee reputation.

A single transaction takes about 5--10 $\mu$s; a typical EVM transaction takes 1--10 ms. That 100$\times$--1000$\times$ gap is what lets us reach 500K+ TPS on commodity hardware.

\subsection{Capability-Weighted Gas}\label{app:capability-weighted-gas}

Gas follows capability, not opcode count. For each INVOKE, we charge the declared capability type, from lightweight query through heavyweight inference, under a base cost class. We also charge payload size, measured as bytes committed to DA, plus composition depth. Each nested INVOKE adds marginal cost against unbounded chains. Reputation discounts still apply: agents with high reputation, low dispute rate, and high completion ratio pay reduced gas, since their histories lower expected dispute-path use. Because sequencing and atomic-execution overhead are shared, we price COMPOSE sub-linearly. A three-hop composition therefore costs less than three independent INVOKEs. Governance controls gas parameters, so rebalancing does not require a hard fork.

\subsection{Atomic Composition Semantics}\label{app:atomic-composition}

When \texttt{\small atomic: true} occurs, a failed constituent INVOKE triggers rollback: if no valid RESPOND arrives within its TTL, the whole composition reverts. We return all escrowed payments and leave reputation unchanged. No partial state transitions are committed. For consistent cross-agent state, we feed each INVOKE's result into later INVOKEs in the dependency graph, so Agent B's output reaches Agent C without a round-trip to the initiator.

A callee can finish its part and still be "clawed back" when a later step fails. Inside a COMPOSE, we hold sub-invocation payments in a \textit{composition escrow sub-account}, rather than releasing them at each RESPOND. Completion only gives the callee a provisional credit. Actual release waits until full COMPOSE success. If COMPOSE reverts because of TTL expiry or an execution error, all provisional credits reverse; the same rule covers out-of-scope CANCEL, and the escrowed funds return to the initiator. This gives the caller zero partial-payment risk and the callee zero partial-delivery risk. At executor level, clawback is atomic: one state-machine transition discards all tentative trie updates and returns all composition escrow balances. During COMPOSE execution, completed sub-invocations keep intermediate results in a non-readable tentative buffer, so third-party trie readers cannot observe them before commit. The window lasts at most the longest constituent INVOKE's TTL plus one block settlement period.

\subsection{Composition Patterns}\label{app:composition-patterns}

Four patterns emerge from the COMPOSE primitive:

A $\rightarrow$ B $\rightarrow$ C gives the simple pipeline case: each agent's output feeds the next agent, and we treat it as the most common pattern. We map it to LangChain/LangGraph chains with on-chain atomicity and payment. In a fan-out trace, A $\rightarrow$ \{B, C, D\}, A invokes multiple agents in parallel and the results are aggregated; within the same batch, the dependency-aware sequencer can run those independent branches concurrently. If A branches as if(result) then B else C, the branch depends on an intermediate result, so the sequencer waits until the conditioning RESPOND is committed. Iterative flows, A $\rightarrow$ B $\rightarrow$ A $\rightarrow$ B, form convergence loops: agents refine the result until a quality threshold is met. The awkward part is termination. We require the executor to enforce max iterations or a convergence criterion to prevent unbounded block consumption.

At the execution layer, we enforce reentrancy protection. Within a COMPOSE, we do not invoke an agent recursively unless the composition graph explicitly permits it with a defined termination condition.

\subsection{Layer Core Ordering Invariant}\label{app:layer-core-ordering}

For every valid block B, the induced transaction DAG G\_B = (V, E) must have the same effect under any topological ordering. Running that ordering through the deterministic transition function T gives an \textit{equivalent} final state: account balances and nonces are identical; escrow and session sequence states are identical too; rate-limit bucket values and reputation deltas match for every agent touched by transactions in B. A COMPOSE transaction is a deterministic state-machine transition over the committed states of its named constituent INVOKEs, and it is enabled only after those constituents have been ordered before it.

During block assembly, a missed write-write conflict can break the invariant unless E records every write-write and read-write conflict among committed transactions. We use the protocol edges for the usual case: RESPOND declares its parent INVOKE, and COMPOSE declares all constituent INVOKEs. Transactions with no path between them must commute under T. When two transactions touch undeclared Layer Core state, such as same-caller balance deductions, escrow status writes, reputation updates, session sequence updates, or rate-limit bucket decrements, we have the sequencer perform \textit{conservative edge-addition}. A lightweight object-ownership index then finds same-object writers and adds an implicit ordering edge.

\textbf{Two-pass construction.} Pass 1 is O(n): walk declared dependencies (RESPOND$\rightarrow$INVOKE edge, COMPOSE$\rightarrow$constituent edges). Pass 2 is O(k), where $k \ll n$ is the number of transactions touching contended Layer Core objects; for each unresolved read-write or write-write pair, add a conservative edge with deterministic direction (earlier nonce wins; lexicographic agent identifier breaks ties). The resulting DAG is complete for Layer Core state: no undeclared conflict enters block execution without an ordering edge. Conservative edges reduce parallelism under same-object contention, but AGNT2's workload model treats agents as the natural sharding boundary, so same-caller contention is expected to be low frequency. The formal serialization-equivalence proof is deferred to \secref{subsubsec:formal-specification}.

\section{Sidecar Failure Handling and Interaction Lifecycle}\label{app:sidecar-failure}

This appendix details deployment models, failure handling, the LLM-heartbeat accommodation, the CPU-overhead-on-GPU analysis, and the per-step interaction lifecycle referenced in \secref{subsec:sidecar-pattern}.

\subsection{Sidecar Architecture (Reference Diagram)}\label{app:sidecar-architecture}

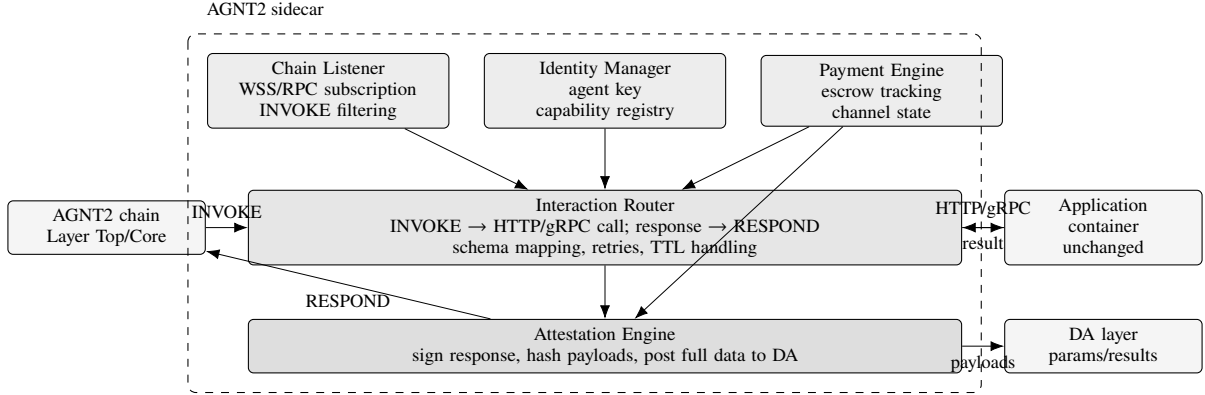
\begin{figure*}[!t]
\centering
\begin{tikzpicture}[
  font=\scriptsize,
  module/.style={draw, rounded corners=2pt, align=center, inner sep=3pt,
                 minimum height=0.54cm, text width=1.18in},
  wide/.style={draw, rounded corners=2pt, align=center, inner sep=3pt,
               minimum height=0.58cm, text width=3.65in},
  external/.style={draw, rounded corners=2pt, align=center, inner sep=3pt,
                   minimum height=0.55cm, text width=0.95in, fill=black!4},
  arrow/.style={-{Latex[length=1.8mm]}, line width=0.35pt},
  dashedbox/.style={draw, dashed, rounded corners=3pt, inner sep=6pt}
]
\node[module, fill=black!7] (chain) {Chain Listener\\WSS/RPC subscription\\INVOKE filtering};
\node[module, fill=black!7, right=0.18in of chain] (identity) {Identity Manager\\agent key\\capability registry};
\node[module, fill=black!7, right=0.18in of identity] (payment) {Payment Engine\\escrow tracking\\channel state};
\node[wide, fill=black!10, below=0.32in of identity] (router) {Interaction Router\\INVOKE $\rightarrow$ HTTP/gRPC call; response $\rightarrow$ RESPOND\\schema mapping, retries, TTL handling};
\node[wide, fill=black!13, below=0.28in of router] (attest) {Attestation Engine\\sign response, hash payloads, post full data to DA};
\node[external, left=0.22in of router] (agnt2) {AGNT2 chain\\Layer Top/Core};
\node[external, right=0.22in of router] (container) {Application\\container\\unchanged};
\node[external, right=0.22in of attest] (da) {DA layer\\params/results};

\draw[arrow] (agnt2) -- node[above] {INVOKE} (router);
\draw[arrow] (router) -- node[above] {HTTP/gRPC} (container);
\draw[arrow] (container) -- node[below] {result} (router);
\draw[arrow] (router) -- (attest);
\draw[arrow] (attest) -- node[below] {RESPOND} (agnt2);
\draw[arrow] (attest) -- node[below] {payloads} (da);
\draw[arrow] (chain) -- (router);
\draw[arrow] (identity) -- (router);
\draw[arrow] (payment) -- (router);
\draw[arrow] (payment) -- (attest);

\draw[dashedbox] ($(chain.north west)+(-0.10in,0.10in)$) rectangle ($(attest.south east)+(0.10in,-0.10in)$);
\node[anchor=west] at ($(chain.north west)+(-0.05in,0.23in)$) {AGNT2 sidecar};
\end{tikzpicture}
\caption{Sidecar module architecture. The application container stays unchanged; we leave chain listening, identity, payment, routing, and attestation to the sidecar.}
\label{fig:sidecar}
\end{figure*}

Figure~\ref{fig:sidecar} details the five module boundaries; the application container is untouched throughout.

\subsection{Deployment Models}\label{app:deployment-models}

Five lines in Docker Compose and the sidecar's running. As a sibling service, it pulls the agent key, capability list, upstream URL, AGNT2 RPC endpoint, and stake registration straight from the env block. The application container doesn't change at all. Kubernetes is more involved: a MutatingWebhookConfiguration injects the sidecar into any pod in a labeled namespace, with capabilities arriving as pod annotations. That's what lets it reach thousands of agents without per-pod work. On bare metal or VM it just runs as a standalone binary instead. Same deal, though. The sidecar's still the only AGNT2 integration point either way.

\subsection{Failure Handling}\label{app:failure-handling}

The sidecar distinguishes three classes of container failure:

A container that OOM-kills looks identical to one that SIGSEGVs or exits cleanly. The sidecar can't tell. Within the liveness probe interval (default 1 s), it catches the container going dark and buffers pending INVOKEs locally, so no RESPOND escapes. Once the TTL deadline passes, the Layer Core sequencer auto-generates a TIMEOUT. The caller gets refunded. Reputation counters on the callee side take the hit. That threshold defaults to 3$\times$ declared max latency, though it's configurable. Exceed it and the sidecar submits a DEREGISTER and pulls the agent from the capability registry.

After ChannelMessage, nothing comes through. The peer sidecar reads that as a crashed Layer Top channel. It won't sit and wait. \textit{T\_heartbeat} = 3 $\times$ 1 s = 3 s; once that window closes without traffic, the counterpart's marked unreachable, and CLOSE\_CHANNEL goes out carrying the latest co-signed state. The channel closes unilaterally. The 4-hour Type 2 challenge window starts from that point. Any ChannelMessages sent but not acknowledged are treated as never sent, so their balances aren't updated. Pending INVOKEs still blocked on a Layer Top RESPOND get promoted to Layer Core, where a fresh on-chain transaction is submitted with a fresh TTL. The original \texttt{\small interaction\_id} still handles deduplication, same as before. Crash-time interactions clear through Layer Core or expire once a TTL-triggered escrow refund fires.

The 3-second default suits callees with low-variance response times. LLM-backed callees are a documented exception. In production, LLM inference latency typically has a long tail: P50 of 200 to 800 ms, P99 at 5 to 20 s for medium-context completions, longer still for chain-of-thought or tool-using agents. Against an LLM peer, a 3-second heartbeat would flag legitimate inference as a crash. Two protocol-level accommodations handle this. First, the heartbeat's a separate channel keepalive, distinct from the inference response. The LLM peer's sidecar emits a heartbeat ack within milliseconds of the previous heartbeat, independent of inference completion. It confirms sidecar liveness but doesn't claim interaction completion. So the 3-second threshold catches sidecar-host crashes (network partition, process death), not slow inference. Second, the per-interaction TTL is the right timer for inference. Set by the caller at INVOKE time. For LLM-class capabilities, callers SHOULD set TTLs aligned with declared P99. The registry stores declared P99 latency per capability, and the caller's sidecar checks TTL $\geq$ 1.5$\times$ declared P99 with warning. Operators with extreme tails (P99 > 30 s, long-running agentic chains) MAY tune \textit{T\_heartbeat} upward (10 s, 30 s) per channel at OPEN\_CHANNEL. It's a per-channel parameter. No global protocol change required.

The awkward case is a late response: the container may still be running, and it can miss TTL. For that case, we let the sidecar count elapsed wall-clock time from INVOKE receipt. With 200 ms left before the TTL deadline, if the local HTTP/gRPC call has not returned, the sidecar cancels the call, submits a RESPOND with \texttt{\small status: TIMEOUT\_GRACEFUL}, and takes the associated reputation penalty. This stops a transient overload from being treated as the harsher escrow-refund path for malice. Operators may also set a \texttt{\small max\_inference\_ms} threshold for LLM containers whose P99 exceeds the interaction TTL. Once set, the sidecar rejects new INVOKEs instead of accepting them and then timing out.

A container can return a syntactically valid response whose semantic content is wrong. In that Byzantine failure case, the sidecar cannot detect the error; we sign and submit the RESPOND exactly as presented. We therefore keep detection of semantic Byzantine failures outside this paper's protocol scope. The attestation + DA record makes the behavior observable and auditable to any third party off-chain, but on-chain adjudication of semantic correctness requires a governance/arbitration layer, which we defer to follow-on work (\secref{subsubsec:qualitative-adjudication}). Callers requiring higher in-band assurance can include a verification agent in their COMPOSE graph.

\subsection{LLM Non-Determinism and the Limits of Container Re-Execution}\label{app:llm-nondeterminism}

A byte mismatch is the artifact that exposes the problem. At temperature > 0, LLM-backed containers are non-deterministic by design, so the \textit{deterministic} container condition behind Off-chain container re-execution no longer holds: identical inputs must give identical outputs. A re-execution may still fit the prompt. But it can differ at byte level, so the comparison fails. For LLM containers, we mark this off-chain audit path as \textbf{unavailable}. By re-execution alone, an auditor can't prove cryptographically that the attested output was wrong. In our current scope, deterrence comes only from the economic and reputational structure, with reputation decay and interaction history cost carrying that role against systematic LLM output fraud. A future qualitative adjudication mechanism (\secref{subsubsec:qualitative-adjudication}) could admit corroborating evidence from lower-temperature re-runs and from third-party verification-agent results into an arbitration record. Its design is left out of scope here.

\subsection{CPU Overhead on GPU-Hosting Machines}\label{app:cpu-overhead}

In a typical interaction, the sidecar's visible work is just WebSocket frame parsing before HTTP dispatch, followed by response hashing and transaction signing. We run it as a standard CPU process. On GPU inference hosts (H100/A100), tokenization and KV-cache management already load-bearing the CPU socket. Batch scheduling reaches that socket too, and NCCL coordination does as well. We measure fewer than 1 ms of CPU compute time per interaction, exclusive of DA I/O wait. That wait adds 10 to 100 ms, depending on DA node proximity, at up to \textasciitilde{}10K concurrent inbound INVOKEs/sec. Small, but not zero. Relative to inference P99 latency, it's still a cost when operators have tight CPU budgets. The sidecar should sit in a dedicated \texttt{\small cpuset} cgroup with NUMA-aware placement, away from GPU kernel dispatch paths.

\subsection{Interaction Lifecycle (per-step)}\label{app:interaction-lifecycle}

\begin{figure*}[t]
\begin{Verbatim}[fontsize=\scriptsize,baselinestretch=0.88,frame=single,framesep=3pt]
TIME    AGENT A (caller)              AGNT2 CHAIN                AGENT B (callee)
-----   ------------------            -----------                ------------------
t=0     A wants to call B's
        "analyze" capability
t=1     A's sidecar:
        1. Looks up B in registry
        2. Posts full params to DA
        3. Submits INVOKE tx --------> INVOKE included in block N
                                       Payment escrowed
t=2                                   Block N propagates
t=3                                   v--------------------> B's sidecar:
                                                              1. Sees INVOKE in block N
                                                              2. Fetches params from DA
                                                              3. Validates escrow
                                                              4. Calls local container
t=4                                                           Container processes
t=5                                                           5. Gets response
                                                              6. Hashes result
                                                              7. Posts result to DA
                                      <-------------------- 8. Submits RESPOND tx
                                       RESPOND in block N+k
                                       Escrow released to B
t=6     A's sidecar:    <-------------+
        1. Sees RESPOND
        2. Fetches result from DA
        3. Delivers to A's container
t=7     A has the result.
        Paid 0.001 AGNT.
        Fully verified on-chain.
\end{Verbatim}
\end{figure*}

\subsection{DA-Post-Failure Recovery}\label{app:da-post-failure}

AGNT2 makes DA publication recoverable. We treat it as a sidecar operation, not as an implicit result of local execution. Each sidecar keeps a durable \textit{transactional outbox} backed by a write-ahead log (WAL). Before the Interaction Router forwards an INVOKE to the container, we persist \texttt{\small \{interaction\_id, caller, callee, params\_hash, received\_at, status=PENDING\}}. Once the container returns, we update the record atomically to \texttt{\small \{status=EXECUTED, result\_hash, executed\_at\}}. Before DA publication is attempted, the sidecar advances it to \texttt{\small DA\_PENDING}. A confirmed DA write then records \texttt{\small DA\_COMMITTED} with \texttt{\small da\_commitment=\{namespace, height, tx\_hash\}}. Only after that does the sidecar submit RESPOND; success moves the record to \texttt{\small RESPONDED}.

DA posting can fail, or just time out. When this happens, our outbox worker retries with exponential backoff, making three attempts at 2s, 4s, and 8s. After the third failure, we take one of two paths. We mark the record \texttt{\small FAILED} and emit a local operator alert; the sidecar does not submit a RESPOND because \secref{subsec:layer-root} rejects state roots whose RESPOND records lack a valid DA commitment (see the T3 discussion in \secref{subsec:representative-threats}). Since RESPOND is idempotent on \texttt{\small interaction\_id}, duplicate submissions are no-ops at Layer Core. The recovery protocol is design-only. Our current prototype DA writer is a mock stub.

\subsection{Malformed LLM Output Handling}\label{app:malformed-output}

Malformed JSON, schema mismatch, hallucinated tool fields, missing required fields, and partial streaming output count as validation failures, not Byzantine behavior, under AGNT2. Only after a container returns HTTP 200 does the Attestation Engine enter a deterministic \textit{schema-validation state machine}. We parse the returned bytes, then check them against the declared output contract, either a JSON Schema or typed capability manifest. If parsing and validation succeed, the sidecar computes \texttt{\small result\_hash} and moves through DA publication and RESPOND. If validation fails, it does not attest to the value.

After a container call, validation may fail; recovery stays local and bounded. We let the Interaction Router retry the same INVOKE parameters up to two times, which means we require that container call to be idempotent for the interaction. If both retries fail validation as well, the sidecar submits RESPOND with \texttt{\small status=PARSE\_FAILURE}, \texttt{\small result\_hash=null}, and \texttt{\small payment=0}. We do not release payment. Escrow goes back to the caller. We record the case in reputation as a soft-failure increment, separate from the Byzantine dispute counter. After three consecutive \texttt{\small PARSE\_FAILURE} responses, we apply a reputation penalty equivalent to a timeout. The caller can still open a quality dispute for an output that is schema-valid but semantically incorrect. \texttt{\small PARSE\_FAILURE} is not a Byzantine event: it triggers no stake slashing and no fraud proof.

\subsection{Concurrent Channel-Open Pre-Finality}\label{app:concurrent-open-channel}

A sidecar cannot exchange Layer Top \texttt{\small ChannelMessage}s just because local acceptance has occurred. Only after Layer Core finality is visible locally do we open the Layer Top channel, and we compute the canonical channel identifier as \texttt{\small channel\_id = H(sort(agent\_id\_A, agent\_id\_B) || first\_accepted\_block\_height || nonce)}. For both peers, \texttt{\small OPEN\_CHANNEL} finality on Layer Core means inclusion in a committed block, not mempool admission. Until that point, our Payment Engine leaves outbound channel messages in its local queue. We do not forward them to the peer.

Two concurrent \texttt{\small OPEN\_CHANNEL} transactions can still name the same ordered agent pair. When that collision appears, we leave the choice to Layer Core and its chain order. The transaction accepted first, by nonce and fee under the chain ordering rules, becomes the canonical channel; the later one reverts. The losing sidecar reads the \texttt{\small REVERTED} status through the Chain Listener and retries with a fresh nonce. After finality holds, both sidecars derive the same \texttt{\small channel\_id} locally. We then release queued messages under that identifier. Any pre-finality message carrying a non-canonical channel identifier is discarded. We do not patch it in place; we resubmit the affected interaction through Layer Core.

\section{Layer Root Fraud Proof Derivations}\label{app:fraud-proof-derivations}

The 1-hour Type 1 challenge window comes first: we derive it, state the VM determinism assumption, trace DA read patterns, and record L1 integration modes.

\subsection{Type 1 Challenge-Window Derivation}\label{app:type1-window}

Under partial synchrony, we derive the 1-hour Type 1 window from the worst-case honest-monitor detection-and-proof cycle, which has four sequential phases.

\begin{enumerate}
  \item \textbf{T\_DA}: retrieving the challenged batch pre-state from the DA layer --- 1--5 minutes under current EigenDA, accounting for geographic distance. Type 1 batch pre-state retrieval is a sequential blob fetch, so DA read amplification is minimal.
  \item \textbf{T\_detect}: computing the correct state root by re-executing the batch and comparing --- under 1 minute, dominated by sequential trie execution of at most 250K typed operations.
  \item \textbf{T\_proof}: constructing the fraud proof transaction that references pre-state and disputing batch --- under 1 minute.
  \item \textbf{T\_submit}: submitting the proof transaction and achieving Layer Root inclusion --- under 5 minutes at expected congestion.

\end{enumerate}
The worst honest case still fits: T\_DA + T\_detect + T\_proof + T\_submit stays inside our budget of about \textbf{8--12 minutes}. The 1-hour window is therefore conservative, with a 5--7$\times$ safety margin for DA outages, including partition recovery and monitor restart. If BFT finality runs on a Layer Root-native L1, T\_submit drops to seconds, and we could shrink the window to 10--15 minutes. The awkward part is still the 4-hour Type 2 channel-dispute window. Both channel participants' sidecars can be offline at once, so recovery must cover container restart, peer reconnection, and a watchdog sweep.

\subsection{VM Determinism Assumption}\label{app:vm-determinism}

A submitted batch manifest must let the fraud-proof verifier replay a canonical serial execution order; for Type 1 fraud-proof re-execution, that means the agent-typed VM produces bit-identical state transitions under bit-identical inputs, across every parallel batch scheduling the dependency analyzer may emit. This is a \textit{required property} of the Core executor, not an implementation detail. We require every opcode in the typed VM to be pure: no wall-clock reads, no thread-local randomness, and no uninitialized memory. For INVOKE/RESPOND/COMPOSE/CHECKPOINT, we can enforce this directly because they operate over trie state and committed payload hashes. Future extensions, including user-defined capability logic, must preserve the same property or stay outside fraud-provable state transitions. Container-side non-determinism (e.g., LLM inference at temperature > 0) belongs in the DA payload, not the VM transition; we handle it at the sidecar/attestation boundary (\secref{subsec:sidecar-pattern} / \appref{app:llm-nondeterminism}), not the fraud-proof boundary.

\subsection{DA Read Patterns}\label{app:da-read-patterns}

\secref{subsec:layer-root} / \secref{subsec:da-bandwidth-gap} measures DA throughput at the sequencer, where the write-bandwidth limit shows up directly. Within our fraud-proof scope, Type 1 + Type 2, DA reads are mostly sequential: Type 1 retrieves a full batch pre-state, and Type 2 reads a channel's signed message history in order. We do not require random-access retrieval of individual interactions. That matters. The 10--30$\times$ shard-reconstruction amplification tied to per-interaction qualitative-fraud review does not appear here. At least near term, the binding DA constraint is the write-bandwidth ceiling, not read amplification. We treat the purpose-built agent-optimized DA layer as a critical follow-on (\secref{subsubsec:scalability-extensions}): regional sharding follows geographic locality, while retention windows follow challenge periods. If future work reintroduces per-interaction qualitative adjudication, the random-access pattern comes back. Near-term deployments may operate at 10K--100K TPS, where existing DA write bandwidth is tractable.

\subsection{Governance Scope}\label{app:governance-scope}

Stake minimums are not set in this paper. Nor are challenge window durations, capability gas weights, fee rates, or DA endpoint configuration. For a production deployment, we still need a governance mechanism for updating these values. We leave that design to follow-on work (\secref{subsubsec:qualitative-adjudication}): constitutional/operational parameter separation, adjudication procedures, and legislative workflows. Here, we treat the parameters as deployment-time configuration.

\subsection{Fraud-Proof Bisection Protocol}\label{app:fraud-proof-bisection}

When a challenger sees $T(s_i, tx) \neq s_{i+1}$, a Layer Core fraud dispute starts by contesting a RESPOND and claiming that the advertised transition is invalid. For AGNT2, we use an interactive \textit{bisection game} to cut the disputed block execution down to one AGNT2 VM step. The construction follows Arbitrum-style interactive fraud proofs. Here, though, execution ranges over AGNT2 VM steps rather than EVM opcodes, and challenger and defender commit to intermediate state roots until only one step remains disputed.

The final proof object contains \texttt{\small \{disputed\_step\_index, pre\_state\_root, post\_state\_root, AGNT2\_step\_witness\}}. We include the interaction trie path, escrow state, and the reputation counter value we need to evaluate the step. Checked against the claimed pre- and post-state roots, an on-chain \texttt{\small AGNT2StepVerifier} contract would verify one VM step; this contract is design-only and not implemented in the current prototype (\secref{subsec:prototype-scope}). This terminal-step check is O(1). But interactive narrowing still needs O(log n) bisection rounds for an n-step disputed execution.

Once a disputed block is larger than 4 MB, it still does not move straight into bisection. AGNT2 makes data availability the gate: DA sampling first has to show that the block data is available, and without a DA availability proof the bisection game is not opened. The testnet challenge window is 24 hours. Under L1 congestion and partial DA outage, production parameters need adversarial modeling, so we defer them to future work. We treat this protocol as design-only and do not implement it in the current prototype.

\textbf{Terminal proof object comparison.} Terminal fraud-proof objects differ mainly in what the verifier has to see. We compare EVM-hosted AGNT2 (Orbit/OP Stack) against native AGNT2 VM; sources are marked [design], [measured†], [published‡], or [estimated§].

\begin{itemize}
  \item \textit{Disputed object}: EVM storage slot update inside Solidity escrow contract [design] vs. typed \texttt{\small RESPOND} trie node update [design].
  \item \textit{Bisection domain}: WAVM / fault-dispute EVM opcode trace vs. AGNT2 VM typed-step trace [design].
  \item \textit{Terminal proof bytes}: full EVM memory+storage snapshot + calldata + stack ($\geq$ 2--5 KB) vs. trie path + escrow state + reputation delta ($\approx$ 1.1 KB) [measured†].
  \item \textit{On-chain verifier work}: re-execute one WAVM/cannon opcode + check EVM state Merkle proof vs. check one typed step against pre/post roots [design].
  \item \textit{Verifier gas}: \textasciitilde{}200K--500K gas per terminal step [published‡] vs. \textasciitilde{}30K--80K gas per terminal step [estimated§].

\end{itemize}
For the native path, we derive the 1.1 KB witness estimate from the \secref{subsec:escrow-gas} measured escrow contract. In an N=10 workflow, 32 trie fields at 32 bytes each give 1,024 bytes; escrow state adds 3 fields $\times$ 32 bytes = 96 bytes; the reputation delta adds 4 counters $\times$ 8 bytes = 32 bytes; total $\approx$ 1,152 bytes $\approx$ 1.1 KB. The EVM lower bound of 2 KB covers the minimum EVM memory page (256 bytes) and storage proof; including full calldata and stack reaches 5 KB, giving a \textbf{1.7--4.3$\times$ witness reduction} (2--5 KB EVM vs. 1.1 KB native) and an estimated \textbf{4--6$\times$ verifier gas reduction}.

†Witness byte estimate derived from the \secref{subsec:escrow-gas} measured contract layout. ‡Arbitrum fault dispute game terminal step gas cost sourced from Arbitrum One public dispute game contracts and Offchain Labs documentation (published, not independently measured here). \S Native AGNT2 verifier gas is a design estimate for an EVM-hosted verifier contract: one Ed25519 signature check (\textasciitilde{}21K gas on EVM) + three 32-byte Merkle path checks (\textasciitilde{}3K gas each) + typed field comparison (\textasciitilde{}5K gas) $\approx$ 35K--80K gas. No \texttt{\small AGNT2StepVerifier} contract exists; this remains a design projection. Building and benchmarking it against the Arbitrum fault-dispute game is follow-on work (\secref{subsubsec:formal-specification}).

\subsection{L1 Integration Modes}\label{app:l1-integration}

Layer Root still works on a generic L1, provided the L1 supports state-root storage and basic contract execution. The L1 can add Layer Root-compatible modules later for the gains we measure. The main one's a native agent-typed VM. It cuts Type 1 re-execution cost by \textasciitilde{}100$\times$ versus EVM, and BFT finality drops to 1--3 s rather than 12+ min, compressing the Type 1 window from 1 hour to minutes. If Layer Root-native L1 isn't available, EigenLayer restaking's another route. An AVS of restaked validators signs pre-confirmations of Layer Root state roots within seconds. That's economic fast finality. No L1 protocol modification needed. For generic settlement, we keep agent-specific logic inside Layer Root and use the L1 only for state-root anchoring and slashing enforcement. Correct, but less efficient. We treat the design of a Layer Root-native L1 as a follow-on contribution (\secref{subsubsec:scalability-extensions}).

\section{Interaction Trie Detailed Schema}\label{app:interaction-trie}

The appendix spells out the full interaction-trie schema. Identity portability and capability discovery are load-bearing --- they're what reputation-as-trie-state, pruning, and the challenge-window retention guarantee from \secref{subsec:layer-core} rest on.

\subsection{Trie Schema}\label{app:trie-schema}

Inspect the AGNT2 L2 artifact directly and you'll find the state root isn't a flat hash at all. It's the Merkle root of the interaction trie. Agent identity anchors the base alongside channel state, with dispute records and interaction logs layered above, capped by the global state entry.

\begin{figure*}[t]
\begin{Verbatim}[fontsize=\scriptsize,baselinestretch=0.88,frame=single,framesep=3pt]
StateRoot
|
|-- /agents/{agent_id}/
|   |-- identity       (name, version, owner, registered_at)
|   |-- capabilities[] (endpoint, input_schema_hash, output_schema_hash, price, avg_latency_ms)
|   |-- economics      (stake, balance, escrowed)
|   |-- reputation     (score, total_completed, total_timeout, total_disputed, disputes_lost, last_active_block)
|   |-- sessions       ({session_id}: SessionState)
|   |-- channels       ({channel_id}: ChannelRef)
|   `-- interaction_history_root: bytes32  // pruned Merkle commitment
|
|-- /channels/{channel_id}/
|   `-- (agent_a, agent_b, deposits, nonce, balances, interaction_count, status, ttl)
|
|-- /interactions/
|   |-- /pending/{interaction_id}
|   |   `-- (invoke_tx, caller, callee, capability, payment, ttl_deadline, composition_id?)
|   |-- /recent/{interaction_id}    // last N blocks, hot storage
|   |   `-- (invoke_tx, respond_tx, caller, callee, capability, payment, execution_ms, status)
|   `-- /history_root: bytes32      // pruned commitment
|
|-- /disputes/{dispute_id}/
|   `-- (target, disputant, dispute_type, status, evidence_hash, opened_at)
|
`-- /global/
    `-- (total_agents, total_interactions, total_channels, total_volume, current_epoch, fee_parameters)
\end{Verbatim}
\end{figure*}

\subsection{First-Class Agent Identity and Portability}\label{app:identity-portability}

Pull the state for any agent in first-class and the nodes resolve to that agent, not the containing object. Under \texttt{\small /agents/\{agent\_id\}/}, identity's rooted in the interaction trie itself, so it's readable and provable on-chain. REGISTER or UPDATE\_CAPABILITIES push updates in; DEREGISTER handles the rest.

\textbf{Identity portability.} Agent identity is anchored on the L1 layer. An agent registered on one AGNT2 L2 instance can be discovered and invoked from another, with reputation and interaction history propagating through L1 settlement --- allowing operators to scale across multiple L2 shards without fragmenting on-chain standing. The identity schema is designed to be compatible with the proposed ERC-8004 on-chain agent identity standard. ERC-8004 is an early-stage proposal (not yet accepted as an EIP draft as of April 2026) that defines a minimal on-chain identity interface for autonomous agents; AGNT2's trie schema is a strict superset of ERC-8004's proposed required fields, enabling interoperability with other ERC-8004-aware contracts and cross-chain agent registries if the standard is adopted. Any ERC-8004-compliant identity could be imported without modification, and any AGNT2 agent identity could be exported to an ERC-8004-compatible registry by projecting the required fields. If the agent-identity standard landscape evolves before Phase 1 testnet launch, the schema will track the prevailing standard.

\subsection{Reputation as Trie State}\label{app:reputation-trie}

At every state transition, the executor updates the trie counters \texttt{\small total\_completed}, \texttt{\small total\_timeout}, \texttt{\small total\_disputed}, \texttt{\small disputes\_lost}, and \texttt{\small last\_active\_block}. We feed these counters into the reputation formula. For a given agent, the score is still deterministic: it remains a function of the agent's interaction trie node, derivable from on-chain data, and verifiable by any participant. Because the formula parameters come from on-chain governance, the agent economy can alter its trust model without hard forks. Dormant agents lose ground. When activity stops, reputation decays, so inactive identities cannot keep high scores indefinitely. Sybil resistance mostly means costs that fake identities cannot share: a minimum stake requirement and a sufficient interaction history. That cost profile makes large-scale reputation farming economically prohibitive.

\subsection{Capability Discovery}\label{app:capability-discovery}

Each agent's trie node carries the on-chain capability registry. When DISCOVER runs, we query that registry by capability name, input/output schema hash, price range, and minimum reputation score. Price, reputation, and latency then decide among agents offering equivalent capabilities. The result is an autonomous market, with no off-chain coordination. Through DISCOVER, a first-class VM primitive, agents find service providers and contract with them entirely on-chain.

\subsection{State Pruning and Challenge-Window Retention Guarantee}\label{app:state-pruning}

At 500K+ interactions per second, full retention stops being practical, so in AGNT2 we use tiered pruning:

For blocks $0$ to $N-K$, the trie still keeps full interaction data in \texttt{\small /interactions/recent/}. From $N-K$ to $N-2K$, we leave a single Merkle root over the batch in the trie and prune the individual records there; the same records remain on DA. Once past $N-2K$, we keep only the commitment root in the trie. The full data stays on DA when dispute material is needed.

We set K so that \texttt{\small /interactions/recent/} holds \textasciitilde{}10 minutes of data. This window covers most dispute and audit use cases , without unbounded growth.

\textbf{Challenge-window retention guarantee.} Trie nodes for interactions under an open dispute are exempt from pruning: the protocol tags each \texttt{\small /interactions/pending/} entry with a \texttt{\small dispute\_hold} flag when a DISPUTE references it; the flag clears only when the dispute closes. For non-disputed interactions, Merkle-root commitments at \texttt{\small /interactions/history\_root} satisfy proof requirements for fraud challenges arising after records leave the hot tier --- provided full payloads remain retrievable on DA, which retains data for at least the longest challenge window (4 hours for Type 2; the 1-hour Type 1 is a strict subset). The invariant is: \textbf{no trie node is pruned while its corresponding DA payload is still within an active or potential challenge window.} The hot tier holds approximately 360 GB of trie state at 500K TPS with a 10-minute window. The 4-hour DA retention requirement at the longest current challenge window is 500K TPS $\times$ 2 KB $\times$ 4 hours $\approx$ 14 TB of DA storage per cycle --- within the purpose-built DA layer's target envelope (\secref{subsubsec:scalability-extensions}). Future re-introduction of qualitative disputes with longer windows would extend this requirement.

\section{Throughput Analysis Detail}\label{app:throughput-analysis}

Per-component $\mu$s/tx figures break down the single-threaded execution ceiling. We also include the hot-agent Zipf analysis with cache-footprint detail, plus the block-broadcast bandwidth derivation and DA-gap prose cited in \secref{subsec:da-bandwidth-gap}.

\subsection{Single-Threaded Execution Ceiling --- Component Breakdown}\label{app:component-breakdown}

Signature verification and trie I/O dominate every typed state transition. The 5--10 $\mu$s/tx single-threaded figure comes from published microbenchmarks and our envelope analysis. Not from a running prototype. We haven't measured it directly.

The per-transaction cost envelope in the table pins the 5-10 $\mu$s single-threaded execution estimate on hardware support. Ed25519 signature verification costs 1.5 to 2.5 $\mu$s, assuming hardware-accelerated AVX-512 or a dedicated crypto unit. Software-only Ed25519 is \textasciitilde{}50 $\mu$s, so hardware acceleration is the load-bearing assumption behind the lower bound. Trie node lookup for the caller \texttt{\small /agents/\{caller\}/} costs 0.5 to 1.0 $\mu$s under an in-memory trie, with a hot working set in L3 cache for the active set (E.2). DRAM-speed lookup in the cold-tail case adds \textasciitilde{}100 ns. Trie node lookup for the callee \texttt{\small /agents/\{callee\}/} runs in the same 0.5 to 1.0 $\mu$s range. Hot callees are the binding case, and cache footprint is analyzed under hot-agent skew below. The interaction record write, \texttt{\small /interactions/pending/}, costs 0.5 to 1.0 $\mu$s, covering append to a ring-buffered pending list plus a Merkle-path update. The amortized cost is dominated by the path-update hash chain (Blake3 / Keccak). The escrow update covers caller balance and callee pending escrow: two trie writes at 0.3 to 0.6 $\mu$s total, both within the agent's own node, so locality's good. TTL/expiry bookkeeping costs 0.2 to 0.4 $\mu$s for deadline-heap insertion, with constant-time amortized behavior. Block manifest append and signature aggregation contribute 0.5 to 1.0 $\mu$s per tx, hashed over the block. That puts the total at 4.0 to 7.5 $\mu$s as a lower bound, or 5.0 to 10.0 $\mu$s with engineering overhead. The 5 to 10 $\mu$s band quoted here includes a \textasciitilde{}25 to 35\% overhead for instrumentation and lock contention on the shared block manifest. It also includes per-tx scheduling cost. The hardware-accelerated Ed25519 entry is load-bearing. If Ed25519 runs software-only, the per-tx budget moves out to 50 to 55 $\mu$s, and the single-threaded ceiling falls to \textasciitilde{}20K TPS. So we treat hardware acceleration as a deployment requirement. It isn't optional. For a production AGNT2 sequencer, we assume a CPU with AVX-512 or comparable signature-verification acceleration. With that, the single-threaded executor saturates at approximately 100K to 200K TPS.

\subsection{Hot-Agent Write Conflicts --- Zipf and Cache Footprint}\label{app:hot-agent-conflicts}

Agent traffic is lopsided. A few high-demand agents, such as a popular pricing oracle or a widely invoked translation service, can take a disproportionate share of INVOKE transactions. Because all INVOKEs aimed at the same callee still write to the same trie node, they serialize. Under heavy workload skew, with 30\% of interactions routed to five hot agents, effective parallelism drops; execution throughput falls to approximately 200K--300K TPS. At the sequencer, we use agent sharding: we partition the interaction pool by callee and give high-frequency destinations dedicated execution lanes with early queuing. With Zipf-distributed service popularity, consistent with microservice traffic benchmarks, the degraded throughput band remains above 300K TPS.

Hot agents hit the same trie nodes again and again. For each hot-agent trie node, we keep identity and reputation state, plus economics and open interaction state; that gives 1--4 KB per agent. For the five busiest agents, the active working set is only 5--20 KB. It still fits inside L3 cache on commodity server hardware. In the cold tail, agents receiving under 10\% of total volume under Zipf pay DRAM-speed trie lookups and add negligible latency to aggregate throughput.

\subsection{Block Broadcast Bandwidth}\label{app:block-broadcast}

At 500K TPS and an average transaction size of approximately 200 bytes (on-chain fields only), a 500 ms block carries 250K transactions $\times$ 200 bytes = 50 MB. That is the hard part. To propagate a 50 MB block every 500 ms, the sequencer needs 800 Mbps of sustained outbound bandwidth. This fits within a single-datacenter NIC, but we cannot use it with a geographically distributed validator set. In production cloud deployments (AWS, GCP, Azure), inter-datacenter links operate at 100--400 Mbps in practice, so full 500K TPS blocks cannot be broadcast to wide-area replicas without compression. In Phase 1, we must keep the full-block validator set within a single datacenter, or connect it by dedicated 10 Gbps links. Geographic distribution waits for Phase 3 ZK compression, which reduces on-chain transaction size by 10--100$\times$ and lowers broadcast bandwidth in the same proportion. Until then, light clients in other regions receive compressed block summaries (state root + interaction count + fee total) at negligible bandwidth, while full block data goes only to co-located validators.

\subsection{The DA Gap (Detail)}\label{app:da-gap-detail}

Execution-layer bottlenecks already cap the design at 300K--500K TPS. In production, though, data availability throughput sets the operational ceiling. At 500K TPS $\times$ 2 KB average payload, the DA layer must absorb approximately 1 GB/sec. Current production DA layers do not get close to that rate. Celestia's current sustained envelope is approximately 6 MB/sec aggregate at the current blob size cap (Table~\ref{tab:da-ceilings}). EigenDA's published mainnet figures, under its current hot-tier configuration, sit in the low tens of MB/sec; the exact figure depends on operator-set size and congestion, so we use \textasciitilde{}10 MB/sec as a conservative headline comparison. This is not a single-digit measurement error. It is a \textasciitilde{}100$\times$ order-of-magnitude mismatch. Higher EigenDA targets, 20--100 MB/sec, come from roadmap and restaked-operator projections, so we treat them as forward-looking bounds rather than current sustained capacity. Near-term AGNT2 deployments at 10K TPS are tractable under existing DA sustained throughput (Table~\ref{tab:da-regimes}); the 50K--100K TPS tier requires near-future DA capacity, and reaching the full execution ceiling requires the purpose-built agent-optimized DA layer identified in \secref{subsubsec:scalability-extensions}.

\subsection{DA Saturation Ceilings and Deployment Regimes}\label{app:da-saturation}

Table~\ref{tab:da-ceilings} reports DA saturation ceilings by payload size [analytical]. Celestia and EigenDA figures come from published sustained bandwidth envelopes; Agent-DA remains a design target.

\begin{table*}[!t]
\centering
\footnotesize
\caption{DA saturation ceilings by payload size per workflow [analytical].}
\label{tab:da-ceilings}
\begin{adjustbox}{max width=\textwidth}
\begin{tabular}{@{}lrrrr@{}}
\toprule
DA system & Bandwidth & Max TPS (0.5 kB/wf) & Max TPS (1 kB/wf) & Max TPS (5 kB/wf) \\
\midrule
Celestia & 6 MB/s & 12,288 & 6,144 & 1,229 \\
EigenDA & 10 MB/s & 20,480 & 10,240 & 2,048 \\
Agent-DA & 200 GB/s & 409.6M & 204.8M & 40.96M \\
\bottomrule
\end{tabular}
\end{adjustbox}
\end{table*}

Table~\ref{tab:da-regimes} gives DA deployment regimes under the canonical payload model (1 KB/interaction, 2$\times$ write amplification). The 65\% Layer Top offload assumption holds for established channel pairs. Tails still matter, though: p99 LLM outputs reach \textasciitilde{}100 KB [analytical].

\begin{table*}[!t]
\centering
\scriptsize
\caption{DA deployment regimes under the canonical 1 KB/interaction, 2$\times$ write-amplification model, 65\% Layer Top offload [analytical].}
\label{tab:da-regimes}
\begin{tabularx}{\textwidth}{@{}rrrrYYY@{}}
\toprule
TPS & Median payload (KB) & Write amp. & Raw DA demand (MB/s) & Effective DA demand (MB/s, 65\% offload) & Candidate DA backend & First-binding bottleneck \\
\midrule
10K & 1 & 2$\times$ & 20 & 7 & EigenDA mainnet & DA write bandwidth \\
50K & 1 & 2$\times$ & 100 & 35 & Near-future DA & DA write bandwidth \\
100K & 1 & 2$\times$ & 200 & 70 & Near-future DA & DA write / block broadcast \\
300K & 1 & 2$\times$ & 600 & 210 & Dedicated DA layer & Block broadcast \\
500K & 1 & 2$\times$ & 1,000 & 350 & Dedicated DA layer & Execution throughput / sequencer CPU \\
\bottomrule
\end{tabularx}
\end{table*}

\subsection{Calldata Field-Width Derivation}\label{app:calldata-derivation}

The linear calldata model in \secref{subsec:da-bandwidth-gap} (\texttt{\small createWorkflow = 164 + 64×N bytes}, \texttt{\small settle = 100 + 256×N bytes}) derives directly from the Solidity ABI encoding of the \secref{subsec:escrow-gas} escrow contract. The fixed term in \texttt{\small createWorkflow} (164 bytes) covers the 4-byte function selector, workflow ID, initiator address, and escrow token fields; the per-agent term (64 bytes) covers one \texttt{\small agent\_id} (32 bytes) and one payment amount (32 bytes). For \texttt{\small settle}, the fixed term (100 bytes) covers the workflow ID and settlement timestamp, with the root hash also encoded here. The per-agent term (256 bytes) covers each agent's result hash (32 bytes) and signature (65 bytes); output metadata accounts for the remaining 159 bytes. These field widths match the OLS regression intercepts and slopes from \secref{subsec:escrow-gas} to within $\leq$3\% across all six measured data points, confirming the model isn't a post-hoc fit but a direct reading of the Solidity ABI layout.

\subsection{Throughput Regime Classification}\label{app:throughput-regimes}

Across all four regimes AGNT2 primitives don't change. Only DA capacity and executor parallelism shift.

\textbf{Sub-DA-bottleneck (10K--100K Layer Core TPS, \textasciitilde{}10M combined with Layer Top)} sits within existing EigenDA/Celestia sustained bandwidth. It's the Near-term deployment target. Parallel-execution (200K--500K TPS, \textasciitilde{}50M combined) requires hardware-accelerated Ed25519 plus multi-worker trie sharding. DA demand exceeds current backends, so it needs the agent-DA layer (\secref{subsubsec:scalability-extensions}). DA-extended (1M--5M TPS, \textasciitilde{}500M combined) requires regional DA sharding at 200 GB/s, and executor parallelism follows from multi-shard trie partitioning. Multi-shard (5M--10M+ TPS, \textasciitilde{}1B+ combined) uses multiple independent Layer Core shards with cross-shard COMPOSE bridging. Follow-on work (\secref{subsubsec:scalability-extensions}).

The first three regimes are within this paper's architecture scope. Multi-shard is deferred.

\section{Extended Threat Model}\label{app:extended-threat-model}

\secref{subsec:representative-threats} summarizes three representative threats; here we give the full adversary-mechanism analyses.

\subsection{T1 --- Byzantine Sequencer: Front-Running and Targeted Re-Ordering}\label{app:threat-front-running}

A concrete failure is easy to state: a Layer Core sequencer sees a pending caller whose INVOKE will move a price-sensitive callee's quoted price, and it inserts its own INVOKE first. The sequencer is controlled by an entity with economic stake in specific interaction outcomes. It observes all pending INVOKEs before a block is sealed, can delay, reorder, or suppress transactions during the submission-to-inclusion window, and can inject its own signed INVOKEs into a block. It cannot forge counterparty signatures. It also cannot modify state transitions without producing an incorrect state root, which Type 1 fraud proofs detect within the 1-hour challenge window. We use three protections against this ordering attack. INVOKEs enter an encrypted mempool under a \textit{threshold-encryption committee} key and are decrypted only after the block ordering has been committed; before that point, the sequencer sees ciphertexts rather than call contents. The committee has \textit{n} nodes, with \textit{n} = 16--64 in our analysis, drawn from a stake-weighted operator set distinct from the sequencer's stake set. They run a threshold-decryption scheme, for example threshold ElGamal or a practical instantiation such as Ferveo, where any \textit{t} members with \textit{t} = $\lceil$2n/3$\rceil$ can jointly decrypt a ciphertext, while no subset of size <\textit{t} can. In block production, the sequencer first commits to the ciphertext-ordered batch and posts that commitment. Only after the commitment is recorded does the committee run threshold decryption; the plaintexts are then executed in the committed order. Committee membership rotates per epoch, as a Layer Root governance parameter deferred to \secref{subsubsec:qualitative-adjudication} for mechanism design. During an epoch, we make members slashable for equivocation, meaning two distinct decryption shares for the same ciphertext, and for liveness failure, meaning no share before the decryption deadline. This adds one off-critical-path round, about \textasciitilde{}100 ms for share aggregation in a regional deployment, amortized over the block interval and not extending per-interaction latency observable to agents. For callers that require bid-protection, we also use commit-reveal for price-sensitive capabilities: the caller posts a commitment in block N and reveals in block N+k, so pre-insertion front-running is infeasible even if mempool contents are visible. And if the sequencer delays a INVOKE past TTL, force-inclusion through L1 lets it be re-submitted to the L1 chain, where Layer Root guarantees inclusion within one settlement window, 30 s--10 min. The residual assumption is committee integrity and L1 availability for the force-inclusion fallback: at least \textit{t} = $\lceil$2n/3$\rceil$ of \textit{n} honest members for decryption, and at most $\textit{t}-1$ colluding members for confidentiality. This is the standard 2/3-honest-majority assumption, structurally identical to the BFT assumption underlying Tendermint, HotStuff, and threshold-encryption literature; AGNT2 inherits that threat model and does not weaken it. If $\geq$ \textit{t} committee members fully collude, the defense falls back to force-inclusion only, giving inclusion but not within-block ordering guarantees. If $\geq$ $\textit{t}-1$ members collude, confidentiality fails before integrity: pending INVOKEs are visible to colluders before sealing. This has a real cost, though it is bounded, and it argues for a larger committee size when MEV exposure is high. Deployments that choose a \textit{single}-operator sequencer with no encrypted mempool bypass the committee scheme entirely. In that mode, AGNT2 still keeps correctness guarantees, so state-root fraud is off the table, but it offers no MEV protection beyond L1 force-inclusion. State-root fraud is off the table in every configuration: the sequencer can re-order, but it cannot rewrite outcomes without detection.

\subsection{T2 --- Colluding Agents: Reputation Manufacture via Synthetic Interactions}\label{app:threat-collusion}

N $\geq$ 2 agents under one principal can try the plain reputation attack by trading synthetic interactions and lifting one or more members above the minimum-reputation threshold that legitimate third-party callers use when choosing a callee. The adversary can open Layer Top channels inside that set, so long as it posts the minimum per-agent stake deposit. Up to deposit capacity, it can also create unlimited signed INVOKE/RESPOND pairs among the colluders and schedule them to resemble legitimate traffic distributions. The awkward part is where the attack stops. It cannot forge signatures for agents outside the colluding set, cannot write directly to the interaction trie, and cannot get a reputation-weighted position without paying the per-identity stake cost. We limit the score gain through the reputation formula rather than by assuming honest traffic. As specified by the schema in \appref{app:reputation-trie}, our formula weights interaction count by the number of distinct counterparties in the agent's history; once the diversity term saturates, synthetic interactions among N colluders add at most log-linear growth in score, no matter how many interactions they generate. Each agent identity also carries a minimum on-chain stake, so manufacturing K distinct synthetic counterparties locks K $\times$ stake\_min in capital. We flag Zipf-violating interaction distributions, including uniform or bimodal counterparty frequencies, for elevated scrutiny and freeze the reputation pending review. Two parameters matter here: the diversity weight must dominate raw volume, with specific values deferred to \secref{subsubsec:cryptoeconomic-design}'s reputation mechanism-design analysis, and stake\_min must be calibrated above the per-interaction reputation yield. If the adversary has state-actor-scale capital, this economic defense weakens. Then anomaly detection is the last line. We state the residual risk plainly: the current design gives economic and statistical deterrence, not a cryptographic guarantee.

\subsection{T3 --- Malicious Sidecar: False DA Attestation of DA-Stored Result}\label{app:threat-malicious-sidecar}

A registered sidecar may hold the agent's signing key and still deliver nothing. It can sign arbitrary attestations, post arbitrary content, or no content, to the DA layer under any commitment hash, and forge local timestamp and execution-duration fields. It cannot forge the caller's signature, and it cannot bypass the protocol-level DA-commitment check applied at RESPOND validation. The attack is to submit a RESPOND whose \texttt{\small result\_hash} field claims a valid deliverable, release the escrow, and never post a payload that a third party could retrieve and verify. We block that path at several points. At RESPOND ingestion, the Layer Core executor applies the protocol-level DA-commitment pre-check check: the \texttt{\small da\_pointer} must reference a DA blob whose advertised content hash matches the attestation's \texttt{\small result\_hash} field. A mismatch rejects the RESPOND, and the INVOKE expires to TIMEOUT with escrow refund. At settlement, Layer Root enforces the DA availability invariant by rejecting any state root that carries RESPOND records whose DA commitments fail a sample-retrieval check. So an `\texttt{\small unreachable'' blob cannot still pay the sidecar. In a COMPOSE graph, downstream agents also fetch and hash-check the payload before committing dependent RESPONDs; if the payload mismatches, they trigger a CANCEL before escrow release. And repeated RESPONDs that fail DA-commitment checks accumulate TIMEOUT penalties on TIMEOUT, pushing the sidecar's reputation below the invocation-eligibility threshold. The residual assumption is DA-layer liveness within the bounded retrieval window, using the Celestia/EigenDA standard of an honest majority of DA operators. If the DA layer fails for operational reasons , the interaction becomes }`posted but unavailable'' and is handled by TTL timeout. The guarantee is narrower: the malicious sidecar cannot silently pocket payment, not that every interaction succeeds under DA failure.

\section{Extended Related Work}\label{app:extended-related-work}

The per-project paragraphs and adjacent-systems engagement are compressed in \secref{subsec:closest-precursors} and \secref{subsec:why-onchain}; we expand them here.

\subsection{Detailed Per-Project Analyses}\label{app:detailed-project-analyses}

Autonolas (Olas) is a Off-chain agent coordination framework with a on-chain service registry. Agents coordinate through an off-chain multi-agent system, and the blockchain records service registration while providing economic incentives. But service invocation and result stay off-chain and unverifiable.

Fetch.ai uses a custom chain (now Cosmos-based), agent messaging, and an agent marketplace. Among the surveyed systems, it sits closest to AGNT2's architectural goals. But it predates the LLM-agent paradigm, and its design targets IoT plus search-and-discovery patterns, not arbitrary capability invocations.

Ritual Network brings ML model execution into the verifiable computation layer by enabling on-chain AI inference. It addresses verifiable ML outputs, not the interaction and coordination problem that AGNT2 targets.

Phala Network uses TEEs to run off-chain agent computation with on-chain verification of execution integrity. The gain is hardware-rooted confidentiality, which AGNT2's attestation model does not require. But Phala does not provide a general-purpose interaction layer for multi-agent coordination.

Hyperledger Fabric is a permissioned-blockchain framework for cross-organizational coordination, with extensive use in supply chain and trade finance consortia, and also in healthcare consortia. Fabric gives each organization certificate authorities for on-chain identity. It also supports private channels for pairwise data segregation and endorsement policies that set per-chaincode quorum rules among participating orgs. We could build an "agent marketplace on Fabric" by treating each agent as a principal under an organization's CA, treating service invocations as chaincode calls with endorsement policies, and using private channels for bilateral relationships. But the failure is at entry, not execution: Fabric does not provide \textit{permissionless} participation. It assumes that a known set of organizations has already ratified a consortium membership policy, so an unknown agent cannot join without bilateral onboarding with existing members. The agent economy AGNT2 targets is permissionless at the protocol level: any Docker container with a key pair and minimum stake can register. So the "why not Fabric?" answer is narrow. Fabric fits consortium coordination with a governance boundary; AGNT2 targets the case where there is no consortium.

An agent marketplace zone fits the Cosmos application-specific pattern: one sovereign chain per application, with IBC handling cross-zone messaging. Fetch.ai (Cosmos-based) remains the closest existing instantiation. Cosmos zones give developers full control over VM, consensus, and governance, and IBC gives a path to cross-zone agent composition. The awkward part is the fit. AGNT2's dependency-aware sequencing, agent-typed VM, and interaction-trie state model together define a specific execution-layer design. We would need custom Cosmos SDK modules for that design, which mostly means building AGNT2 on Cosmos rather than as a rollup on EVM L1. That is a valid engineering route, and v2 could target Cosmos as the host. The research contribution, the execution-layer design itself, does not depend on the substrate. The other difference is cost. AGNT2's fraud-proof model follows the optimistic-rollup lineage rather than Tendermint BFT: the 1-hour challenge window amortizes cost across many batches, while Tendermint BFT pays per-block finality cost. At 500K TPS, the gap compounds materially.

\subsection{ROCOCO/Vegeta Detailed Positioning}\label{app:vegeta-positioning}

ROCOCO and Vegeta are the nearest mechanistic precedents for AGNT2's Dependency Analyzer, so we must position the paper's contribution against both at three levels.

ROCOCO targets general-purpose distributed transactions, where eager locking and two-phase commit add coordination before work can run. Instead, ROCOCO records declared dependencies at submission time. Once predecessors have completed, it replays transactions in dependency order. The main observation in ROCOCO is that dependency information from the transaction initiator can replace runtime conflict detection, which matters directly for AGNT2. Vegeta takes dependency-graph execution into parallel smart-contract execution for leaderless blockchains: it speculatively executes transactions, builds a conflict graph from observed read-write set intersections, and re-executes in topological order. Against DeFi workloads, Vegeta reports 7.8$\times$ speedup over serial EVM execution.

Against AGNT2, the narrowing is easiest to see. AGNT2's Dependency Analyzer still belongs to the same \textit{mechanistic} lineage, but we should not read it as a cheaper answer to the same task. It answers a \textbf{different problem}. ROCOCO and Vegeta handle general-purpose concurrent transactions, where dependencies are read-write storage conflicts. Since the protocol does not already reveal the workload's dependency structure, those conflicts have to be found at runtime. That is the hard part. Their contribution is correct parallel execution under this condition. AGNT2 instead works on typed agent service invocations, a narrower workload. Here, protocol semantics already carry the dependencies: every RESPOND names its parent INVOKE by hash, and every COMPOSE names its constituent INVOKEs.

\begin{enumerate}
  \item \textbf{O(n) construction instead of O(n²) conflict detection.} Each transaction contributes exactly one edge per declared dependency; no pairwise intersection computation is needed. This is a direct consequence of the domain narrowing, not an algorithmic improvement over Vegeta --- it is infeasible in Vegeta's setting because EVM storage access patterns are not declared at submission.
  \item \textbf{Correctness semantics, not performance optimization.} The INVOKE-before-RESPOND invariant is part of the interaction protocol's correctness definition; violating it is a protocol fault. In ROCOCO/Vegeta's setting, dependency-order violation is a serializability violation --- a performance/consistency issue, but the transactions are still "valid" computations under EVM semantics. AGNT2's dependencies are \textit{semantic}, not merely ordering hints.
  \item \textbf{Trust domain.} ROCOCO targets symmetric-trust database participants; Vegeta targets permissioned BFT validators; AGNT2 targets mutually-untrusting agents across organizational boundaries. The declared-dependency approach in AGNT2 relies on the \textit{callee's} signature on a RESPOND naming its parent INVOKE --- the dependency edge is cryptographically attested by a mutually-untrusting counterparty, not by a client in a trusted database cluster.

\end{enumerate}
AGNT2 should not be read as "Vegeta plus agents": we are not describing a better algorithm for the same problem. We treat it as the minimal DAG-execution design that follows from a typed agent-service-invocation protocol, and that shifts the research question and the tractability profile.

\subsection{Rotating TEE Committee --- Extended Engagement}\label{app:tee-committee}

A quorum-signed log update from N independent TEE-running attestors, rotated each epoch, only approximates the append-only integrity properties of an on-chain state root. The rotating cross-organizational TEE committee is still the strongest off-chain substitute we've got for an on-chain reputation system. But two practical obstacles remain.

A missing TEE attestation after committee majority collusion leaves a minority principal with only the committee's own arbitration. There isn't another remedy. That follows from treating the committee, at the collective level, as a trusted aggregator. So reputation derivability still depends on an honest committee quorum, which in this setting is typically the 2/3 assumption. On-chain reputation is different: a party reading the state root can derive it while assuming only L1 consensus. Committees can give integrity. But not \textit{aggregator-free} integrity. The guarantee is the committee's, not the protocol's.

At the first counterparty in a new ecosystem, the handshake can stop. The agent needs to present reputation history, yet that party doesn't know the agent's old TEE committee. An on-chain trie node makes the evidence travel with the agent instead. The Merkle path verifies against a known state root. A committee-signed log is messier: the verifier still has to check the committee's legitimacy, then follow its rotation history. And when committees shift across ecosystems and epochs, our on-chain approach avoids that extra trust-transfer work.

At the point where the two obstacles coincide, the failure becomes visible. Even the strongest rotating TEE committee gives us committee-level integrity. It doesn't give protocol-level aggregator-free integrity, and portability remains committee-bounded rather than self-verifying. We close that gap with the on-chain trie root.

\section{Evaluation Claims and Requirements}\label{app:evaluation-claims}

\subsection{Claim-to-Evidence Map (C1--C10)}\label{app:claim-to-evidence}

The claim-to-evidence map links AGNT2's ten primary evaluation claims to Evidence levels. These are Measured (prototype), Simulated, Analytical, and Design-only. The O(n) Dependency Analyzer ordering claim, C1, has Simulated evidence with Medium strength in \secref{subsec:dependency-analyzer}. It would be upgraded by running the analyzer on production agent traces with adversarial dependency patterns. The 1000$\times$ parallelism claim on the mixed-10K workload is also Simulated, Medium strength, in \secref{subsec:dependency-analyzer}, and it needs replay of real multi-agent transaction traces at larger scale. The Sequencer DAG block production claim, C3, has Simulated, Medium evidence in \secref{subsec:dependency-analyzer}. A distributed sequencer cluster under fault injection would strengthen it. The DA bandwidth gap of \textasciitilde{}100$\times$ at 500K TPS, C4, is Analytical with Medium--High strength in \secref{subsec:da-bandwidth-gap}, pending validation against measured DA throughput on candidate backends. For zero-code-change sidecar integration, C5 is Measured, Medium strength, in \secref{subsec:zero-code-sidecar}. It needs integration of multiple unmodified agent frameworks and reported compatibility failures. The channel open/close latency below 100ms result is Measured, Medium strength, in \secref{subsec:channel-latency}, but it should be measured under WAN conditions with concurrent channel churn. The escrow gas result, 244K/638K/1,096K at N=1/10/20, is Measured with High strength, in \secref{subsec:escrow-gas}. It should be repeated across additional EVM networks and contract compiler versions. For COMPOSE atomic settlement and rollback, C8 is Measured, Medium strength, in \secref{subsec:compose-settlement}. Rollback still needs Byzantine counterparty and sequencer faults. The Layer Core 300K--500K TPS execution ceiling, C9, remains Design-only, Low strength, in \secref{subsec:layer-core-execution} until an end-to-end Layer Core prototype is built and benchmarked. The C10 unilateral-close + lockout enforcement claim is Measured, Medium strength, in \secref{subsec:sidecar-crash-recovery} (SQLite-backed crash recovery remains follow-on; long-duration recovery tests under injected crashes and partitions are the upgrade path). Requirements Satisfaction for R1--R7 is covered in \appref{app:requirements-satisfaction}.

\subsection{Requirements Satisfaction (R1--R7)}\label{app:requirements-satisfaction}

Established pairs hit <100ms P2P through Layer Top. Layer Core handles new interactions and compositions in 500ms--2s, with the sidecar auto-routing traffic to the minimum-overhead layer. That's the concrete path by which R1 satisfies the Sub-second interaction latency requirement in the AGNT2 design. Rich payload support for R2 works by keeping full payloads in the DA layer, posting them to DA. We place the 32-byte commitment hash on-chain on chain; structured capability schemas stay in interaction-trie nodes. Payload bytes avoid EVM calldata cost.

R3 handles Dependency-aware transaction ordering through the Dependency Analyzer and Batch Builder, pulling DAG from the pending pool first: INVOKEs always precede their RESPONDs, though independent INVOKEs still run in parallel within the same block. Identity's a different story. In R4, agent identity is treated as first-class state --- anchored in the interaction trie and the proposed ERC-8004 path. Each \texttt{\small /agents/\{agent\_id\}/} trie node carries identity, capabilities, economics, reputation, sessions, and channels. And that's where ERC-8004 compatibility for cross-chain portability lives.

For micropayments, R5 uses Layer Top state channels with OPEN\_CHANNEL and per-message signed balance updates, applying CHECKPOINT compression to settle up to 1,000 interactions per on-chain transaction. Composable multi-agent atomic transactions in R6 use the COMPOSE primitive: an ordered dependency graph of INVOKEs executes atomically, and any TTL miss or failure reverts the whole composition with a full escrow refund. R7 splits computational fraud proof coverage between Type 1 and Type 2. Type 1 covers agent-typed VM for computational disputes; re-execution sits in the same 1-hour window. Type 2 covers nonce-ordering and double-signing slash with a 4-hour window.

\subsection{Safety Claim Taxonomy}\label{app:safety-taxonomy}

Table~\ref{tab:safety} maps each safety property to its adversary model, evidence type, and residual risk.

\begin{table*}[!t]
\centering
\footnotesize
\caption{Safety claim taxonomy. Each protocol safety property mapped to adversary model, evidence type, and residual risk.}
\label{tab:safety}
\begin{tabularx}{\textwidth}{@{}YYlYY@{}}
\toprule
Safety property & Adversary model & Evidence type & Artifact / basis & Residual risk \\
\midrule
Escrow atomicity: no payment without valid result & Malicious callee & Measured & Foundry test suite T4--T9: invalid-sig revert, window expiry, nonce enforcement (\secref{subsec:compose-settlement}) & Byzantine counterparty fault injection not yet tested \\
DA integrity: no RESPOND without retrievable payload & Malicious sidecar withholding DA & Measured & Layer Root rejects state roots without valid DA commitment; T3 analysis in \secref{subsec:representative-threats} & DA-layer liveness assumed (honest majority of DA operators) \\
Ordering invariant: topological equivalence & Byzantine sequencer reordering & Analytical & Two-pass O(n)+O(k) construction \secref{subsec:layer-core}; simulation confirms batch counts & Formal serialization-equivalence proof deferred (\secref{subsubsec:formal-specification}) \\
Type 1 fraud detectability & Byzantine sequencer fraudulent state root & Design-only & Challenge window derivation \appref{app:type1-window}; bisection protocol \appref{app:fraud-proof-bisection} & No \texttt{\small AGNT2StepVerifier} contract built; fraud-proof VM is design-only \\
Channel dispute resolution: higher-nonce state wins & Stale-state submission & Measured & \texttt{\small initiateUnilateralClose} + 24-hour EVM time-advance test (\secref{subsec:sidecar-crash-recovery}) & Double-signing not exercised under adversarial behavior \\
Sybil resistance & Colluder manufacturing K identities & Analytical & Stake-floor $\times$ K cost + diversity-weighting saturation analysis in \secref{subsec:representative-threats} & Sufficiently capitalized adversary is bounded economically, not cryptographically \\
Sequencer censorship resistance & Byzantine sequencer suppressing txs & Analytical & Force-include via L1 + TTL $\geq$ 2$\times$ settlement interval (\secref{subsec:threat-coverage}) & Phase 1 single-sequencer; decentralized sequencer in Phase 3 \\
Parse-failure isolation & Callee returning schema-invalid output & Design-only & State machine \appref{app:malformed-output}; idempotency contract specified & Schema-valid semantically wrong output (qualitative fraud) deferred to \secref{subsubsec:qualitative-adjudication} \\
Channel-open uniqueness & Concurrent OPEN\_CHANNEL race & Design-only & State machine \appref{app:concurrent-open-channel}; channel\_id = H(sort(agent\_id\_A, agent\_id\_B) ‖ first\_accepted\_block\_height ‖ nonce) & Not tested under adversarial race conditions \\
\bottomrule
\end{tabularx}
\end{table*}

\subsection{Comparative Analysis}\label{app:appendix-comparative-analysis}

It's easier to see in nine dimensions. Table~\ref{tab:comparative} compares AGNT2 against service-mesh, the nearest rollup, and the main agent-blockchain alternatives across all nine dimensions.

\begin{table*}[!t]
\centering
\footnotesize
\caption{Comparative analysis across nine evaluation dimensions.}
\label{tab:comparative}
\begin{tabularx}{\textwidth}{@{}YYYYYY@{}}
\toprule
Dimension & Optimism / Arbitrum & Kubernetes + Istio & Fetch.ai & Autonolas & \textbf{AGNT2} \\
\midrule
Interaction frequency & Seconds & Milliseconds & Seconds & Off-chain & <100ms--2s \\
Payload support & Generic calldata & HTTP body & Message passing & Off-chain & Structured DA-backed \\
Dependency-aware ordering & No & No & No & No & Yes (DAG-based) \\
Agent identity & EOA / contract & Service account & On-chain address & On-chain registry & First-class trie node \\
Per-interaction micropayment & No & No & Limited & No & Yes (escrowed INVOKE) \\
Multi-agent atomic composition & No & No & No & No & Yes (COMPOSE) \\
Computational fraud proof & Yes (EVM, 7-day) & No & No & No & Yes (agent-typed VM, 1-hour) \\
Trust model & Trustless, financial & Trusted operator & Partial & Off-chain trust & Trustless, cross-principal \\
Zero-code agent onboarding & No & Partial & No & No & Yes (sidecar) \\
\bottomrule
\end{tabularx}
\end{table*}

\subsection{Security Coverage}\label{app:security-coverage}

Table~\ref{tab:security} maps each threat class to its coverage mechanism, assessment, and residual risk.

\begin{table*}[!t]
\centering
\footnotesize
\caption{Security coverage by threat class.}
\label{tab:security}
\begin{tabularx}{\textwidth}{@{}YYYY@{}}
\toprule
Threat & Coverage mechanism & Assessment & Residual risk \\
\midrule
Service fraud, deterministic containers & Attestation audit trail + reputation decay & Medium: verifiable off-chain; adjudication deferred & No automated on-chain slashing until governance layer added \\
Service fraud, LLM containers (temp > 0) & Reputation decay only & Low: reputational deterrence only & Qualitative adjudication deferred to \secref{subsubsec:qualitative-adjudication} \\
Collusion rings & Stake cost + diversity weighting + pattern detection & Medium: costly at scale; detectable & Low-volume slow-burn collusion below detection threshold \\
DoS / interaction flooding & Stake-based quotas + sidecar rate limiting & High: attacker bears infrastructure cost & Coordinated multi-account flood requires large stake \\
Data poisoning in compositions & Verification agent + on-chain audit trail & Medium: opt-in, not enforced & Pipelines omitting verification agents remain exposed \\
Sequencer censorship & Force-include via L1 + decentralized sequencer in the phase 3 design & High long-term; medium in the phase 1 design & The phase 1 single-sequencer design requires L1 fallback \\
Agent MEV & Encrypted mempool + commit-reveal & Medium: partial mitigation & Full MEV resistance requires decentralized sequencer \\
Sequencer downtime & Layer Top fallback + L1 force-include & High: agents with channels unaffected & Agents without channels blocked until sequencer recovers \\
Sybil attacks & Stake requirement + interaction history cost & High: economic cost of per-identity staking & Economically bounded, not cryptographically \\
\bottomrule
\end{tabularx}
\end{table*}

Formal analysis hasn't run on these yet. Rough benchmarks, for now. Community testing will shift some before they're locked.

\end{document}